\documentclass[usegraphicx,usenatbib,useAMS]{mn2e}

\usepackage{multirow}
\usepackage{xcolor}
\usepackage{aas_macros}
\usepackage{amssymb}
\usepackage{xspace}
\usepackage[normalem]{ulem}
\usepackage{enumitem}

\newcommand{\gsim}{\gtrsim} 
\newcommand{\lsim}{\lesssim} 

\newcommand{\kms}{\,{\rm km}\,{\rm s}^{-1}}
\newcommand{\Msol}{\,{\rm M}_{\odot}}

\newcommand{\Mpc}{{\rm Mpc}}

\newcommand{\GALFORM}{\textsc{galform}\xspace}
\newcommand{\lgalaxy}{\textsc{l-galaxies}\xspace}
\newcommand{\MORGANA}{\textsc{morgana}\xspace}

\newcommand{\AREPO}{\textsc{arepo}\xspace}
\newcommand{\SUBFIND}{\textsc{subfind}\xspace}

\title[Cooling model comparison]
{A comparison between semi-analytical gas cooling models and 
cosmological hydrodynamical simulations}

\author[Hou et al.]  {Jun Hou\thanks{\scriptsize E-mail:
    jun.hou@durham.ac.uk (JH);cedric.lacey@durham.ac.uk (GCL)},$^{1,2}$ Cedric G. Lacey\footnotemark[1],$^2$
  Carlos. S. Frenk$^2$
  \\
  $^{1}$Shanghai Key Lab for Astrophysics, Shanghai Normal University, 
  100 Guilin Road, Shanghai 200234, China
  \\
  $^{2}$Institute for Computational Cosmology, Department of Physics,
  University of Durham, South Road, Durham, DH1 3LE, UK}

\begin{document}

\maketitle

\begin{abstract}We compare the mass cooling rates and cumulative
  cooled-down masses predicted by several semi-analytical (SA) cooling
  models with cosmological hydrodynamical simulations performed using
  the \AREPO code (ignoring processes such as feedback and chemical enrichment). The SA
  cooling models are the new \GALFORM cooling model introduced in
  \cite{new_cool}, along with two earlier \GALFORM cooling models and
  the \lgalaxy and \MORGANA cooling models. We find that the
  predictions of the new \GALFORM cooling model are generally in best agreement
  with the simulations. For halos with $M_{\rm halo}\lesssim 3\times
  10^{11}\Msol$, the SA models predict that the timescale for
  radiative cooling is shorter than or comparable to the gravitational
  infall timescale. Even though SA models assume that gas falls onto
  galaxies from a spherical gas halo, while the simulations show that
  the cold gas is accreted through filaments, both methods predict
  similar mass cooling rates, because in both cases the gas accretion
  occurs on similar timescales. For halos with $M_{\rm halo}\gtrsim
  10^{12}\Msol$, gas in the simulations typically cools from a roughly
  spherical hot gas halo, as assumed in the SA models, but the halo
  gas gradually contracts during cooling, leading to compressional
  heating. SA models ignore this heating, and so overestimate mass
  cooling rates by factors of a few. At low redshifts halo major
  mergers or a sequence of successive smaller mergers are seen in the
  simulations to strongly heat the halo gas and suppress cooling,
  while mergers at high redshifts do not suppress cooling, because the
  gas filaments are difficult to heat up. The new SA cooling model
  best captures these effects. \end{abstract}


\section{Introduction}\label{sec:introduction}
An important goal of cosmology is to understand the physical origin of various galaxy properties and how they evolve over cosmic time. However, this task has not been completed because it involves a large number of complex physical processes that take place on a wide range of spatial and temporal scales. Currently, there are two main theoretical methods to study galaxy formation and evolution, hydrodynamical simulations \citep[e.g.][]{illustris_simulation,eagle_simulation} and semi-analytical (SA) models \citep[e.g.][]{white_rees_1978,WF1991,baugh_2006_SA_review,benson_2010_SA_review}. The former attempt to solve numerically the hydrodynamical equations relevant to galaxy formation, and can provide many detailed predictions, but at a high computational cost, and it remains challenging to generate large galaxy samples for statistical studies. In contrast, SA models focus mainly on global properties of galaxies, such as the total stellar mass and total cold gas mass of a galaxy, treat the physical processes driving the evolution of these properties as channels connecting mass reservoirs, and model these processes with highly simplified prescriptions. By doing so, SA models significantly reduce the computational cost and complexity, and it is easy to generate large galaxy samples, but they provide less detailed predictions for galaxy properties. Because of their versatility and low computational cost, SA models can be used to experiment with varying model assumptions and parameters, an important methodology which is much more difficult with hydrodynamical simulations \citep{eagle_parameter_search}. In reality, the two methods are complementary, and combining them is the best stratagem to gain better understanding of galaxy formation and evolution. For this purpose, it is important that SA models should be as physically based as possible in order to provide real physical insight into galaxy formation.

Among all of the physical processes involved in galaxy formation, heating of gas in shocks, followed by radiative cooling and subsequent accretion onto galaxies are of crucial ones, because they determine the total amount gas in a galaxy that is available for subsequent processes such as star formation, black hole growth and gas ejection by feedback. This process can be well resolved in current hydrodynamical simulations, which provide detailed predictions for gas accretion rates onto galaxies. On the other hand, the treatment of gas cooling in SA models is  based on significant simplifying assumptions, for example, that the gas cools from a spherical hot gas halo. Even with these assumptions, a proper calculation of gas cooling would still need to trace the thermal history of each spherical gas shell, but doing this directly, which in principle requires a 1D hydrodynamical simulation, is too complex for a SA model, and runs counter to the advantages such models gain from their simplicity. Furthermore, a gas shell could take a long time to cool down, during which time the hot gas halo may evolve significantly, and this further complicates tracing its thermal history.

A number of different approaches to modelling gas cooling in SA models have been introduced in the literature. The approaches used in earlier versions of the \GALFORM SA model \citep{cole2000,galform_bower2006,benson_bower_2010_cooling} follow the cooling history of the halo gas in a very rough way, while instead the cooling models used in \lgalaxy \citep[e.g.][]{munich_model1} and \MORGANA \citep[e.g.][]{morgana1} do not explicitly follow this history. The new cooling model introduced in \citet{new_cool} improves on the earlier \GALFORM cooling models by introducing a more accurate approximation for the thermal history of the gas shell that cools down at a given timestep. It also follows the evolution of the hot gas halo in more detail, not just including the contraction of the hot halo induced by cooling, but also including the hot halo adjustment induced by dark matter halo growth. The new cooling model should therefore be more physically realistic than the other models mentioned above. Since current hydrodynamical simulations are largely able to resolve the cooling process, they can provide a good test for the modelling of gas cooling in SA models. In this work, we assess the accuracy of our new cooling model, as well as the simplifiying assumptions made in it, by comparing its predictions for mass cooling rates and cumulative cooled down masses with hydrodynamical simulations. We also compare the other cooling models mentioned above with the same simulations.

Similar comparisons of SA models with hydrodynamical simulations have been performed in several previous works \citep[e.g.][]{Benson_2001_comp,Yoshida_2002_comp, Helly_2003_comp,Cattaneo_2007_comp,morgana2,Saro_2010_comp,Lu_2011_comp, Hirschmann_2012_comp, monaco_2014_comp}. All of these works compare SA models and hydrodynamical simulations in a 'stripped down' galaxy formation scenario, in which physical processes such as star formation and feedback are ignored, in order to focus on gas cooling. In early works the comparisons were typically very simple; for example, \citet{Benson_2001_comp} only compared the cooling properties averaged over all halos of a given mass, with the SA model implemented on Monte Carlo halo merger trees instead of trees extracted from the corresponding cosmological dark matter simulation. Over time, these comparisons have become increasingly sophisticated, with the recent work by \citet{monaco_2014_comp} being particularly sophisticated one.

In this work, we follow the approach of \citet{monaco_2014_comp} in the following respects: we use stripped-down hydrodynamical simulations and SA models for our comparison; we limit our comparisons of the gas cooling to only well resolved dark matter haloes (resolved with at least $2000$ particles, but using a different method from that in \citeauthor{monaco_2014_comp}); we run all of the SA models on halo merger trees extracted from N-body simulations carried out using the same initial conditions, same code and same resolution as for the hydrodynamical simulations; and we perform a halo-by-halo comparison instead of comparing only averages over a halo population.

In contrast to previous works, which are mainly based on the SPH
(Smooth Particle Hydrodynamics) method, in this work the simulations
are carried out using the moving-mesh code \AREPO
\citep{arepo}. According to \citet{Nelson2013_hydro_cooling},
simulations using SPH may introduce some artificial effects into the
thermal history of the gas, which is crucial for the cooling
calculation. 

The simplified gas cooling picture contained in SA models can highlight important physics from the rich detail generated by hydrodynamical simulations, and thus allow one to learn more from the simulations. In this work, we try to gain insight into some of the detailed cooling physics through comparisons between the predictions of the new cooling model and simulation results for several individual halos. More specifically, we study the temperature and density distributions of the gas in dark matter haloes that are predicted by SA models to be in either the fast cooling regime (radiative cooling faster than gravitational infall onto central galaxies) or slow cooling regime (cooling slower than infall) respectively, and also the thermal properties of the gas during halo major mergers.  Many previous works have emphasized the gas accretion contributed by cold filaments \citep[e.g.][]{cold_accretion_keres,dekel_filament_accretion,Nelson2013_hydro_cooling}. This filamentary accretion is dominant mainly in haloes in the fast cooling regime. Here we make comparisons for halos in both the fast and slow cooling regimes, in order to derive a more complete view of gas cooling. The effects of halo major mergers on cooling were previously studied in \citet{monaco_2014_comp} based on SPH simulations, while here we study this in more detail and using moving mesh simulations.

This paper is organized as follows. \S\ref{sec:methods} first provides
an introduction to the simulations used in this work, and describes
how the halo merger trees are constructed and how the mass cooling
rates are measured from simulations; then in
\S\ref{sec:method_SA_models} we give a brief description of the SA
cooling models considered in this work. \S\ref{sec:results} sets out
the main results. \S\ref{sec:results_cold_vs_filamentary} investigates
some details of gas cooling in the fast cooling regime, while
\S\ref{sec:results_slow_cooling_regime} investigates details of gas
cooling in the slow cooling regime, and
\S\ref{sec:results_halo_mergers} investigates the effects of halo
major mergers on
cooling. \S\ref{sec:results_model_comp_case_study}
  and \S\ref{sec:stat_comparison} then provide further comparisons
between the simulations and different SA cooling models, in order to
assess the accuracy of each SA model. Finally, our main conclusions
are summarized in \S\ref{sec:summary}.


\section{Methods} \label{sec:methods}
\subsection{Moving mesh code AREPO for hydrodynamics} \label{sec:method_arepo_code}
\AREPO is a finite volume grid-based hydrodynamical code \citep{arepo}. The grid is generated by a Voronoi tessellation of space, where this tessellation is induced by a set of grid generation particles. These particles are allowed to have arbitrary motions, but usually they are set to largely follow the motion of the fluid itself. Then the fluid fluxes across the boundaries of each cell in the grid are calculated using the exact 1D Riemann solution, and these fluxes are used to update the whole fluid field.

This method can be viewed as an amalgam of the adaptive mesh refinement (AMR) approach and smooth particle hydrodynamics (SPH). Compared to the more traditional grid-based AMR method, allowing the grid to move with the fluid has several advantages. Firstly, this can largely avoid large fluid velocities relative to the grid. Large fluid velocities lead to the kinetic energy dominating the total energy budget of the flow, leading to very inaccurate estimation of the internal energy and of the thermal state of the fluid, which is crucial for calculating gas cooling. In cosmic structure formation, gas flows with large relative velocities are common, and this means that large velocities relative to the grid are inevitable for a static grid in the AMR method. Secondly, the moving mesh provides a continuous adjustment of the resolution, instead of the discrete jump of resolution in the mesh refinement of AMR. The latter artificially suppresses structure growth due to gravity \citep[e.g.][]{AMR_gravity_problem1,AMR_gravity_problem2}.

The SPH method is commonly used in cosmological hydrodynamical
simulations. This method is particle-based and quasi-Lagrangian, which
also gives it a continuously adaptive resolution. Continuous fluid
quantities, such as density, are derived though smoothing over nearby
particles. This smoothing introduces relatively large artificial
dissipation and diffusion, which can broaden shock fronts, leading to
less efficient shock heating, and damp turbulent motions, leading to
artificial heating \citep{Bauer_2012}. As shown in
\citet{Nelson2013_hydro_cooling}, these effects bias the cooling
calculation. The grid-based flux calculation in \AREPO largely avoids
these effects, although averaging quantities within
  cells still results in some numerical diffusion
  effects. Corrections can be added to SPH to mitigate those
artificial effects in specific situations
\citep[e.g.][]{improved_SPH}, but the general applicability of these
corrections and potential side effects remain unclear.

In summary, the moving mesh code \AREPO is an ideal tool for the study of gas cooling in the context of cosmic structure formation.


\subsection{Simulations} \label{sec:method_simulations}
We assume the $\Lambda$CDM cosmology with cosmological parameters based on the WMAP-7 data \citep{CMB_obs_WMAP7}: $\Omega_{{\rm m}0}=0.272$, $\Omega_{\Lambda 0}=0.728$, $\Omega_{{\rm b}0}=0.0455$ and $H_{0}=70.4\kms{\Mpc}^{-1}$, and an initial power spectrum with slope $n_{\rm s}=0.967$ and normalization $\sigma_{8}=0.810$. 

We ran simulations in two cubes, with co-moving sizes of $50\Mpc$ and $25\Mpc$ respectively, both of them with periodic boundary conditions. The initial conditions were generated by the code {\textsc{n-genic}\xspace} \citep{N_GenIc}. We used $752^3$ dark matter particles and initially the same number of gas cells in the large cube, and $376^3$ dark matter particles and gas cells in the small cube. For all of our simulations the dark matter particle mass is $9.2\times 10^6\Msol$, and the gravitational softening scale is $0.98$ co-moving ${\rm kpc}$. There are $128$ output times evenly spaced in $\log (1+z)$, from $z=19$ to $z=0$. The physical time interval between two adjacent outputs is about one quarter of the halo dynamical timescale, $t_{\rm dyn}$, which is defined as $t_{\rm dyn}=r_{\rm vir}/V_{\rm vir}$, with $r_{\rm vir}$ and $V_{\rm vir}$ being respectively the virial radius and velocity of a halo.

We ran two dark matter only simulations in the $25\Mpc$ and $50\Mpc$ cubes respectively, to construct merger trees for the semi-analytical (SA) gas cooling models. We also ran an adiabatic gas simulation (i.e.\ without gas cooling or other physical process such as star formation and feedback) in the $25\Mpc$ cube. This was used to investigate the hot gas density and temperature distributions used in the SA models. Finally we ran two gas simulations with cooling, in the $25\Mpc$ and $50\Mpc$ cubes respectively. The simulation with cooling in the small cube has a relatively small data volume and is useful for selecting interesting individual halos for detailed case studies, while the simulation with cooling in the large cube contains enough halos to derive statistical results.

For the simulations with gas cooling, we adopt cooling functions for primordial gas based on \citet{hydro_cooling_function}, but do not include the cooling due to inverse compton scattering on the CMB, which is less important than other cooling mechanisms considered here. There is no UV heating background, but we impose a cooling temperature floor to prevent gas cooling in very small dark matter halos. Specifically, a gas cell can cool only if its temperature $T_{\rm gas}$ satisfies
\begin{equation}
T_{\rm gas}>T_{\rm cool,lim}=3.5\times 10^4\times [\Omega_{{\rm m}0}(1+z)^3+\Omega_{\Lambda 0}]^{1/3}\,{\rm K}, \label{eq:cooling_T_limit}
\end{equation}
where $z$ is the redshift, and $T_{\rm cool,lim}$ roughly corresponds to the virial temperature of a halo with $M_{\rm vir}=2\times 10^{10}\Msol$, which in our simulations is resolved with $2000$ particles. According to \citet{monaco_2014_comp}, this resolution is high enough for reliable cooling calculations. These simulations do not include any feedback or metal enrichment processes.

The gas that has cooled down  would accumulate in the halo centre and reach very high density. This cold and dense gas has a very short dynamical timescale, leading to large computational cost due to the condition on the timestep, but because this gas has already been accreted by the central galaxy, its further fate is irrelevant to the gas cooling calculation. Therefore we turn this gas into collisionless stellar particles to save computation time. As in \citet{monaco_2014_comp}, the gas is turned into stars when its density is higher than $\delta_{\rm sfr,lim}\bar{\rho}_{\rm gas}$ and its temperature is lower than $T_{\rm sfr,lim}$, where $\bar{\rho}_{\rm gas}=\Omega_{\rm b}(z)\rho_{\rm crit}(z)$ is the mean gas density, with $\Omega_{\rm b}(z)$ and $\rho_{\rm crit}(z)$ the baryon fraction and critical density at redshift $z$ respectively, and $\delta_{\rm sfr,lim}$ and $T_{\rm sfr,lim}$ are two parameters. We adopt $\delta_{\rm sfr,lim}=10^4$ and $T_{\rm sfr,lim}=\min[10^5\,{\rm K},T_{\rm cool,lim}]$. Note that here this star formation is not meant to represent a physical process but is just a numerical technique to reduce the computation time.

The structures formed are first identified through the friends-of-friends (FOF) algorithm \citep{FOF_algorithm}, and then each FOF group is further split into subgroups using \SUBFIND \citep{munich_model1}.


\subsection{Merger trees} \label{sec:merger_tree_construction}
The halo merger trees are constructed using the Dhalo algorithm \citep{Dhalo_tree1,Dhalo_tree2}. This method is based on the subgroups identified by \SUBFIND. It links the subgroups at different snapshots by cross matching their most bound dark matter particles to generate the merger trees for these subgroups. These subgroups are then grouped into Dhalos by examining their separations. If one subgroup lies within twice the half mass radius of another subgroup, then they are defined to be in the same Dhalo. Thus a structure and the substructures it contains are assembled into a single Dhalo, while the structures enclosed in a single FOF group through artificial low density bridges are separated into different Dhalos. Once a subgroup belongs to a Dhalo, it is always considered to be part of this Dhalo. This ensures that a subhalo temporarily leaving its host halo during a merger is treated as being a subhalo since its first infall. Finally, the subgroup merger trees are combined to derive the Dhalo merger trees. The mass of a Dhalo is the sum of the masses of all subgroups belonging to it. Subgroup masses are provided by \SUBFIND. For dark matter only simulations, the mass of a given subgroup is the total mass of dark matter particles in this subgroup, while for hydrodynamical simulations, the mass of a subgroup is the total mass of dark matter particles, stellar particles and gas cells that belong to this subgroup. Unless otherwise specified, all the halo masses used in this work are the Dhalo masses.

The Dhalo merger trees of the dark matter only simulation are built for calculating SA models, while the merger trees of the hydrodynamical simulation are built to extract the gas cooling histories from this simulation. The merger trees of these two simulations are linked by cross matching the $50$ most bound dark matter particles of the base halos at $z=0$. Two linked merger trees are treated as being of the same halo in different simulations.


\subsection{Measuring the cooled down gas mass in hydrodynamical simulations} \label{sec:measurement_from_hydro}
The cooled down gas in a halo sinks towards the minimum of its gravitational potential well, and is accreted by the galaxy there. According to the \SUBFIND algorithm, this potential minimum is usually associated with the most massive subgroup in a Dhalo. Thus, the cooled down gas should also be found in the region around the potential minimum of this subgroup. We identify this region as the central galaxy in the Dhalo. Further, as mentioned in \S\ref{sec:method_simulations}, in our simulations the cool gas accreted by galaxies is quickly turned into stars, so in the end the cooled down gas is represented by the stars in the central region of the most massive subgroup of a given Dhalo. For simplicity, here the central region is defined as a sphere of radius $20$ co-moving kpc around the centre. We have checked that our measurements are reasonably stable for different choices of this aperture radius.

The above-mentioned selection defines the stars in the central galaxy of a given Dhalo. However, the gas cooled over the history of this Dhalo along the major branch  of its merger tree (formed by the most massive progenitor Dhalos) only forms part of the stars; the other part is formed in other galaxies and is delivered to the central galaxy through galaxy mergers. The stars from these two channels can be separated based on two features of galaxy mergers. Firstly, the time from the first infall of a satellite galaxy to its final merger with the central galaxy is typically longer than one halo dynamical timescale. Secondly, the gas cooling in a satellite halo is expected not to last for a long time after its infall into the host halo, so when a satellite has nearly merged with the central galaxy, it should contain very few, if not zero, newly formed stars. 

Motivated by these two observations, after we pick out the stars in the central galaxy of a given Dhalo at the $i$-th output time $t_{\rm i}$, we then go back to this halo's main progenitor (defined as the most massive progenitor Dhalo) at the $(i-1)$-th output time $t_{{\rm i}-1}$, and remove all the selected stars that also exist at $t_{{\rm i}-1}$. This should only leave the stars formed by the gas cooled down in the given Dhalo between $t_{{\rm i}-1}$ and $t_{\rm i}$. The reason for this is as follows. The stars in the central galaxy at $t_{\rm i}$ can be divided into three categories, namely the stars in the main progenitor of this central galaxy at $t_{{\rm i}-1}$, the stars delivered by the merging satellites during $(t_{{\rm i}-1}, t_{\rm i}]$ and the stars newly formed in the central galaxy between $t_{{\rm i}-1}$ and $t_{\rm i}$. Because the time interval corresponding to $(t_{{\rm i}-1}, t_{\rm i}]$ is shorter than the halo dynamical timescale, at $t_{{\rm i}-1}$, these merging satellites should be in the current halo's main progenitor halo, so the above method should cover all of the merging satellites, and remove all stars formed before $t_{{\rm i}-1}$ in either the main progenitor of the central galaxy or in these merging satellites, leaving only new stars formed during $(t_{{\rm i}-1}, t_{\rm i}]$. Hence the selected stars are all newly formed within $(t_{{\rm i}-1}, t_{\rm i}]$. As argued above, by the time a satellite has nearly merged with the central galaxy, the gas cooling rate onto the satellite should be very low, so the selected stars should be mainly formed by cooled down gas accreted onto the central galaxy.

With stars selected in this way, the mass of gas cooled down within $(t_{{\rm i}-1}, t_{\rm i}]$, $\Delta M_{\rm cool,i}$, is measured as
\begin{equation}
\Delta M_{\rm cool,i}=\sum_{j=1}^{N}m_{\rm star,j},
\end{equation}
where the index $j$ labels the selected stellar particles, $N$ is their total number, and $m_{\rm star,j}$ is the mass of the $j$-th stellar particle. Then, the gas cooling rate at $t_{\rm i}$ is estimated as
\begin{equation}
\dot{M}_{\rm cool}(t_{\rm i})=\frac{\Delta M_{\rm cool,i}}{t_{\rm i}-t_{{\rm i}-1}}.
\end{equation}
The cumulative cooled down mass, $M_{\rm cool}(<t_{\rm i})$, is calculated as
\begin{equation}
M_{\rm cool}(<t_{\rm i})=\sum_{j=i_{\rm start}}^{i} \Delta M_{\rm cool,j},
\end{equation}
where the summation is along the major branch of a merger tree, (namely it only includes cooling in the main progenitors), and $i_{\rm start}$ is the index of the earliest output time reached by this branch.


\subsection{Semi-analytical Calculation of Gas Cooling} \label{sec:method_SA_models}
\subsubsection{New cooling model (Hou et al. 2017)} \label{sec:method_SA_models_new_cool}
This cooling model is described in detail in \citet{new_cool}. It assumes that the halo gas is initially in a spherical hot gas halo, and gradually cools and falls onto the central galaxy. If the radiative cooling is faster than the infall due to gravity, then the gas becomes cold before it reaches the central galaxy, and this cold gas forms a cold gas halo. Gas cooling reduces the pressure support from the halo centre outwards, leading to the contraction of the hot gas halo. Diffuse gas newly accreted during dark matter halo growth is assumed to be shocked heated to the virial temperature of the halo and joins the existing hot gas halo. The growth of the dark matter halo also induces the adjustment of the hot gas halo.

The new cooling model assumes that the hot gas halo has a single temperature, which is the virial temperature, $T_{\rm vir}$, of the corresponding dark matter halo, and a density profile described by the $\beta$-distribution:
\begin{equation}
\rho_{\rm hot}(r)\propto \frac{1}{r^2+r_{\rm core}^2},\ r_{\rm cool,pre}\leq r \leq r_{\rm vir},
\label{eq:rho_hot_new_cool}
\end{equation}
where $r_{\rm cool,pre}$ is the inner boundary of the hot gas halo, and its calculation will be described later, while $r_{\rm vir}$ is the halo virial radius, and is the outer boundary of the hot gas halo, and $r_{\rm core}$ is the core radius. We calculate $r_{\rm vir}$ from the current halo mass and the current mean halo density according to the spherical collapse model, $\Delta'_{\rm vir} \rho_{\rm crit}$, with  $\rho_{\rm crit}$ being the current critical density of the universe (see Appendix~\ref{app:r_core_hot}). In this work we adopt $r_{\rm core}$ based on the hot gas density profiles measured from the hydrodynamical simulation without cooling, and details are given in Appendix~\ref{app:r_core_hot}. The normalization of the density profile is fixed by requiring the total mass in this profile to equal the total hot gas mass.

The gas cooling is calculated in a sequence of finite timesteps. Within a given timestep $[t, t+\Delta t)$, the hot gas halo is assumed to be static. Gas cooling starts from the halo centre. By the end of the current timestep, $t+\Delta t$, there is an outer boundary, $r_{\rm cool}$, that separates the cooled down and hot gas. The gas shell at $r_{\rm cool}$ has just cooled down by $t+\Delta t$.

$r_{\rm cool}$ is calculated as follows. It is assumed that a gas shell of radius $r$ and infinitesimal thickness $\delta r$ cools down when it has radiated away all of its thermal energy, namely when $\delta U=\delta E_{\rm cool}$, where $\delta U$ is the thermal energy of this shell and $\delta E_{\rm cool}$ is the energy it has lost by cooling radiation. Defining $r_{\rm cool}$ as the radius of the shell that has cooled down at $t+\Delta t$, this condition becomes $\delta U=\delta E_{\rm cool}(t+\Delta t)$. This condition can be further expressed in terms of the so-called cooling timescale, $t_{\rm cool}$
\begin{equation}
t_{\rm cool}(r_{\rm cool})=t_{\rm cool,avail},
\label{eq:r_cool_def}
\end{equation}
where \begin{equation}
t_{\rm cool}(r) \equiv \frac{\delta U}{\delta L_{\rm cool}}=\frac{3k_{\rm B}}{2\mu_{\rm m}}\frac{T_{\rm vir}}{\tilde{\Lambda}(T_{\rm vir})\rho_{\rm hot}(r)},
\label{eq:t_cool_def}
\end{equation}
with $\tilde{\Lambda}(T_{\rm vir}) \rho_{\rm hot}^2$ being the radiative cooling rate per unit volume, $\delta L_{\rm cool}=\tilde{\Lambda}(T_{\rm vir}) \rho_{\rm hot}^2(r) \, 4\pi r^2 \delta r$ being the shell's current cooling luminosity, $k_{\rm B}$ the Boltzmann constant and $\mu_{\rm m}$ the mean molecular mass of the hot gas, and
\begin{equation}
t_{\rm cool,avail} \equiv \frac{\delta E_{\rm cool}(t+\Delta t)}{\delta L_{\rm cool}}
\label{eq:t_cool_avail_def}
\end{equation}
is defined to be the time available for cooling. Although this provides the formal definition of $t_{\rm cool,avail} $, we actually calculate it by an approximate method, as follows. The hot gas halo is assumed to be static within a timestep, so that $\delta E_{\rm cool}(t+\Delta t)=\delta E_{\rm cool}(t)+\delta L_{\rm cool}\Delta t$, and
\begin{eqnarray}
t_{\rm cool,avail} & = & \delta E_{\rm cool}(t)/\delta L_{\rm cool} +\Delta t \nonumber \\
                   & \approx & E_{\rm cool}/L_{\rm cool}+\Delta t,
\label{eq:t_cool_avail_new_cool}
\end{eqnarray}
where $L_{\rm cool}(t)$ is the total cooling luminosity of the current hot gas halo, and $E_{\rm cool}(t)$ is the total energy radiated away by the current hot gas halo up to time $t$. \citet{new_cool} argue that equation~(\ref{eq:t_cool_avail_new_cool}) provides a good approximation to $t_{\rm cool,avail}$ as defined by equation~(\ref{eq:t_cool_avail_def}).

In the above, $L_{\rm cool}(t)$ is calculated as
\begin{equation}
L_{\rm cool}=4\pi\int_{r_{\rm cool,pre}}^{r_{\rm vir}}\tilde{\Lambda}(T_{\rm vir}) \rho_{\rm hot}^2(r)r^2dr ,
\end{equation}
while $E_{\rm cool}$ is calculated using the following recursion relation, starting from the initial value $E_{\rm cool}=0$:
\begin{eqnarray}
E_{\rm cool}(t+\Delta t) & = & E_{\rm cool}(t)+L_{\rm cool}(t)\times \Delta t \nonumber \\
& - & L'_{\rm cool}(t)\times t_{\rm cool,avail},
\label{eq:E_cool_recursive}
\end{eqnarray}
where 
\begin{equation}
L'_{\rm cool}(t)=4\pi\int_{r_{\rm cool,pre}}^{r_{\rm cool}}\tilde{\Lambda}(T_{\rm vir}) \rho_{\rm hot}^2(r)r^2dr.
\end{equation}
The second term in equation~(\ref{eq:E_cool_recursive}) adds the energy radiated away in the current timestep, while the third term removes the contribution to the radiated energy from gas between $r_{\rm cool,pre}$ and $r_{\rm cool}$, because that gas cools down in the current timestep and therefore is not a part of the hot gas halo at the next timestep.

With $r_{\rm cool}$ known, together with the inner boundary of the hot gas halo $r_{\rm cool,pre}$ and the density profile, it is straightforward to derive the mass of gas cooled down in the current timestep (which is the gas between $r_{\rm cool,pre}$ and $r_{\rm cool}$). This mass joins the cold gas halo mass, $M_{\rm halo, cold}$. The gas in the cold gas halo is not pressure supported, and so is assumed to free fall onto the central galaxy. Based on this, the mass, $\Delta M_{\rm acc,gal}$, accreted onto the central galaxy over a timestep is calculated as
\begin{equation}
\Delta M_{\rm acc,gal}=M_{\rm halo,cold}\times \min[1,\Delta t/t_{\rm ff}(r_{\rm cool})],
\label{eq:M_gal_acc_new_cool}
\end{equation}
where $t_{\rm ff}(r)$ is the free-fall timescale at radius $r$. Note that $M_{\rm halo,cold}$ is increased by gas cooling, so on a timescale $t_{\rm cool}(r_{\rm cool})$, and it is depleted on a timescale $t_{\rm ff}(r_{\rm cool})$, so if the cooling is slower than the infall, $t_{\rm cool}(r_{\rm cool})>t_{\rm ff}(r_{\rm cool})$, then $M_{\rm halo,cold}$ remains very small compared to the mass of the hot gas halo.

If the hot gas halo remained static, then the inner boundary of the hot gas halo at next timestep, $r_{\rm cool,pre}(t+\Delta t)$, should just be $r_{\rm cool}(t+\Delta t)$ calculated at the current timestep. However, as mentioned previously, the cooling and dark matter halo growth in the current timestep induces adjustments in the hot gas halo, so some further adjustments of $r_{\rm cool}(t+\Delta t)$ are required. Gas cooling reduces the pressure support of the hot gas halo, causing it to contract. This contraction is driven by gravity, so its effect is modelled as
\begin{equation}
r_{\rm cool,pre}(t+\Delta t)=r_{\rm cool}(t+\Delta t)\times \max[0,1-\Delta t/t_{\rm ff}(r_{\rm cool})].
\label{eq:r_cool_pre_cooling_contraction}
\end{equation}
The above equation is only valid for a static gravitation potential well. If the dark matter halo grows in the current timestep, then the gravitational potential changes. We estimate the effect of this by requiring that the mass of dark matter within $r_{\rm cool,pre}$ remains the same before and after halo growth, i.e.\ 
\begin{equation}
M'_{\rm halo}[r'_{\rm cool,pre}(t+\Delta t)]=M_{\rm halo}[r_{\rm cool,pre}(t+\Delta t)],
\label{eq:r_cool_pre_halo_growth}
\end{equation}
where the primed quantities are after halo growth, and the unprimed quantities are before halo growth. Here the dark matter is used to trace the adjustment, because the gas within $r_{\rm cool,pre}$ is cold with negligible pressure, so it should have similar dynamics to that of the collisionless dark matter. At the starting time, there has been no cooling and all of the halo gas is hot, so $r_{\rm cool,pre}=0$, while equations~(\ref{eq:r_cool_pre_cooling_contraction}) and (\ref{eq:r_cool_pre_halo_growth}) are used to determine $r_{\rm cool,pre}$ for later timesteps.


\subsubsection{\GALFORM cooling model GFC1} \label{sec:method_SA_models_GFC1}
The GFC1 (GalForm Cooling 1) cooling model was first introduced in \citet{galform_bower2006}, and is used in all recent \GALFORM models \citep[e.g.][]{galform_gonzalez2014,galform_lacey2015}. It shares many features of an earlier cooling model introduced in \citet{cole2000}.

Both the GFC1 and \citeauthor{cole2000} cooling models split each branch of a halo merger tree into segments separated by artificial halo formation events. A halo without any progenitor at the previous timestep is flagged as being a halo formation event, and a halo two or more times more massive than the progenitor in the previous halo formation event is flagged as corresponding to a new halo formation event. 

The \citeauthor{cole2000} cooling model then assumes that the hot gas
halo is static between two adjacent halo formation events, and is
reset at each halo formation event. At a halo formation event, the hot
gas is assumed to be newly heated, with  a single temperature, $T_{\rm
  vir}$, and distributed from $r=0$ to $r=r_{\rm vir}$ with the
$\beta$-distribution as its density
profile. $r_{\rm vir}$ and $T_{\rm
  vir}$ are calculated at each halo formation
  event using the mean halo density obtained from the spherical
  collapse model, and are kept constant until the next halo formation
event. The gas accreted between two halo formation events is delayed
from joining the hot gas halo until the next halo formation event, and
so the integral over the hot gas density profile from $r=0$ to
$r=r_{\rm vir}$ is remains constant between formation events, and
equal to $M_{\rm hot}+M_{\rm cooled}$, where $M_{\rm hot}$ is the mass
left in the hot gas halo, and $M_{\rm cooled}$ is the mass of gas
cooled down in this halo since the last halo formation event. This
constant is used to fix the density profile normalization. Since the
gas is newly heated at the halo formation event, and the hot halo is
static, for any shell $\delta E_{\rm cool}(t)=\delta L_{\rm
  cool}(t-t_{\rm form})$, where $\delta L_{\rm cool}$ is the
(constant) cooling luminosity of this shell and $t_{\rm form}$ is the
time of the last halo formation event. Then one has 
\begin{equation}
t_{\rm cool,avail}(t)=\delta E_{\rm cool}(t)/\delta L_{\rm cool}=t-t_{\rm form}.
\label{eq:t_cool_avail_GFC1}
\end{equation}
$r_{\rm cool}$ is then obtained by solving equations~(\ref{eq:t_cool_avail_GFC1}) and (\ref{eq:r_cool_def}).

The GFC1 model largely inherits the above calculation, but with several modifications. Firstly, the virial radius $r_{\rm vir}=GM_{\rm halo}/V_{\rm vir}^2$ is now calculated by using the current halo mass rather than the halo mass at the last halo formation event, so partially including the effect of halo growth. However, $V_{\rm vir}$ here is still the value at the last halo formation event, in order to keep $T_{\rm vir}=\mu_{\rm m}V_{\rm vir}^2/(2k_{\rm B})$ the same as that at the last halo formation event. Secondly, the newly accreted gas joins the hot halo immediately after accretion, so now $M_{\rm hot}$ includes the contribution from this gas.

These two modifications make the hot gas halo not exactly static between halo formation events, so equation~(\ref{eq:t_cool_avail_GFC1}) is not completely justified in the GFC1 model. Also note that although the hot halo is not static, the halo contraction induced by cooling is still largely ignored. To see this, consider the gas cooling in a static dark matter halo. In this case, the GFC1 model reduces to the \citeauthor{cole2000} model, so the hot gas halo is also static. 

In the present paper, $r_{\rm core}$ in the $\beta$-distribution is determined by using the correlation described in Appendix~\ref{app:r_core_hot}. This differs from what was done in earlier applications of the \citeauthor{cole2000} and GFC1 cooling models.

Neither the GFC1 model nor the \citeauthor{cole2000} cooling model
include a cold gas halo. Instead they include the effect of the
gravitational infall timescale for the
cooled down gas through the so called free-fall radius $r_{\rm ff}$,
which is calculated as 
\begin{equation}
t_{\rm ff}(r_{\rm ff})=t_{\rm ff,avail},
\label{eq:r_ff_def}
\end{equation}
where $t_{\rm ff,avail}$ is called the time available for free-fall,  for which these two models adopt $t_{\rm ff,avail}=t-t_{\rm form}$. This is then used to calculate the infall radius $r_{\rm infall}$, defined as
\begin{equation}
r_{\rm infall}=\min[r_{\rm cool}, r_{\rm ff}].
\label{eq:r_infall_def}
\end{equation}
The gas within $r_{\rm infall}$ should have cooled down and fallen onto the central galaxy by the end of the current timestep. The other radius needed is the previous infall radius $r_{\rm infall,pre}$, which is calculated through
\begin{equation}
4\pi\int_{0}^{r_{\rm infall,pre}}\rho_{\rm hot}(r)r^2dr = M_{\rm cooled}.
\label{eq:r_infall_pre_def}
\end{equation}
The gas within this radius should have fallen onto the central galaxy before the current timestep. Thus, the mass of gas accreted onto the central galaxy within the current timestep, $\Delta M_{\rm acc,gal}$, is
\begin{equation}
\Delta M_{\rm acc,gal}=4\pi\int_{r_{\rm infall,pre}}^{r_{\rm infall}} \rho_{\rm hot}(r)r^2dr.
\label{eq:M_gal_acc_GFC1}
\end{equation}
$\Delta M_{\rm acc,gal}$ is also added to $M_{\rm cooled}$ to update it.

Note that if the radiative cooling is faster than the gravitational infall, then $r_{\rm cool}>r_{\rm ff}$, and the cooled down gas between $r_{\rm ff}$ and $r_{\rm cool}$ is left in the hot gas halo and treated as hot gas in the next timestep.


\subsubsection{\GALFORM cooling model GFC2} \label{sec:method_SA_models_GFC2}
The GFC2 (GalForm Cooling 2) model was introduced in
\cite{benson_bower_2010_cooling}. It largely removes the dependence of
the gas cooling on artificial halo formation
events the earlier \GALFORM cooling models. 

This model still assumes that the hot gas halo has a single temperature, $T_{\rm vir}$, which is now the virial temperature of the current dark matter halo, rather than of the halo at the last halo formation event. In the present study, the density profile is again assumed to be the $\beta$-distribution, $\rho_{\rm hot}\propto 1/(r^2+r_{\rm core}^2)$, with $r_{\rm core}$ determined from the correlation described in Appendix~\ref{app:r_core_hot}. In this work, we do not include SN feedback, and in that case the normalization of the density profile is fixed by requiring that the integral of this profile from $r=0$ to $r=r_{\rm vir}$ equals $M_{\rm hot}+M_{\rm cooled}$\footnote{In the case that gas is ejected from the halo by SN feedback, then an extra term is included for the ejected gas; see \citet{benson_bower_2010_cooling} for more details}. Here, $M_{\rm hot}$ is the total hot gas mass in the current halo, just as in the GFC1 model. However, $M_{\rm cooled}$ is different from that in the GFC1 model: (a) it is increased by the mass of gas cooled down and accreted onto the central galaxy; (b) it is gradually reduced according to
\begin{equation}
\dot{M}_{\rm cooled}=-\alpha_{\rm remove}\times M_{\rm cooled}/t_{\rm ff}(r_{\rm vir}),
\label{eq:M_cooled_dot_GFC2}
\end{equation}
with $t_{\rm ff}(r)$ the free-fall timescale at radius $r$ and $\alpha_{\rm remove}\sim 1$ being a free parameter; (c) it is propagated to a halo from its most massive progenitor instead of being reset to zero at each halo formation event as in the GFC1 model. 

The gradual reduction of $M_{\rm cooled}$ described by equation~(\ref{eq:M_cooled_dot_GFC2}) is intended to model the effect of gravitational contraction of the hot gas halo due to loss of pressure support resulting from cooling of gas in the central regions of the halo. It acts to lower the density profile normalization, while leaving $M_{\rm hot}$ unchanged; after the reduction of $M_{\rm cooled}$, the hot gas has to be distributed to smaller radii, or in other words, the hot halo contracts towards the halo centre. This is more physical than what is assumed in the GFC1 model. However, here the contraction happens on a timescale $\sim t_{\rm ff}(r_{\rm vir})$, while the contraction should be most significant in a region of radius $\sim r_{\rm cool}$, so a more realistic timescale should be $\sim t_{\rm ff}(r_{\rm cool})$, as in the new cooling model.

The cooling radius $r_{\rm cool}$ is calculated through equation~(\ref{eq:r_cool_def}), while  $t_{\rm cool,avail}$ is calculated through equation~(\ref{eq:t_cool_avail_new_cool}), but $L_{\rm cool}(t)$ and $E_{\rm cool}(t)$ are calculated differently from the new cooling model. Specifically,
\begin{eqnarray}
L_{\rm cool}(t) & = & 4\pi\int_{0}^{\rm vir}\tilde{\Lambda}(T_{\rm vir}) \rho_{\rm hot}^2(r)r^2dr \nonumber \\
             & \approx & \bar{\rho}_{\rm hot}(t)\times 4\pi\int_{0}^{\rm vir}\tilde{\Lambda}(T_{\rm vir}) \rho_{\rm hot}(r)r^2dr \nonumber \\
             & = & \bar{\rho}_{\rm hot}(t)\tilde{\Lambda}(T_{\rm vir})[M_{\rm hot}(t)+M_{\rm cooled}(t)],
\label{eq:L_cool_GFC2}
\end{eqnarray}
where 
 $\bar{\rho}_{\rm hot}$ is the mean density of the hot gas halo. Approximating $\rho_{\rm hot}^2(r)$ as $\bar{\rho}_{\rm hot}\rho_{\rm hot}(r)$ is very rough, and when the hot gas distribution extends in to near the centre of the halo, which is typical when the cooling is much slower than the infall, $\rho_{\rm hot}(r)>\bar{\rho}_{\rm hot}$,  so this approximation tends to underestimate the cooling luminosity.

$E_{\rm cool}(t)$ is calculated by integrating $L_{\rm cool}$ over time, namely
\begin{equation}
E_{\rm cool}(t)=\int_{t_{\rm init}}^{t}L_{\rm cool}(\tau)d\tau + \int_{t_{\rm init}}^{t}\frac{3k_{\rm B}}{2\mu_{\rm m}}T_{\rm vir}\dot{M}_{\rm cooled}d\tau,
\label{eq:E_cool_GFC2}
\end{equation}
where $t_{\rm init}$ is the initial time, and the second term removes the contribution to $E_{\rm cool}$ from the gas removed from the $M_{\rm cooled}$ reservoir, because this gas is no longer a part of the hot gas halo.

Similar to the GFC1 model, the GFC2 cooling model does not include any
cold gas halo. It still calculates the effect of the gravitational
infall timescale of the cooled down gas through
$r_{\rm ff}$, which is calculated by using
equation~(\ref{eq:r_ff_def}), however, the time available for
free-fall, $t_{\rm ff,avail}$, is calculated differently in the GFC2
model. Specifically, a quantity with dimensions of energy similar to
$E_{\rm cool}$ is cumulated, but this quantity is not allowed to
exceed an upper limit $t_{\rm ff}(r_{\rm vir})\times L_{\rm cool}$,
and $t_{\rm ff,avail}$ is then calculated as the ratio of this
quantity to $L_{\rm cool}$. The upper limit ensures that $t_{\rm
  ff,avail}\leq t_{\rm ff}(r_{\rm vir})$. This procedure for
calculating $t_{\rm ff,avail}$ does not seem physically well
motivated, since the gravitational infall rate should not depend on
the cooling luminosity. 

After calculating $r_{\rm ff}$, the GFC2 model then calculates $r_{\rm infall}$ and $r_{\rm infall,pre}$ by using equations~(\ref{eq:r_infall_def}) and (\ref{eq:r_infall_pre_def}) respectively, as in the GFC1 model. The mass accreted onto the central galaxy is then calculated by using equation~(\ref{eq:M_gal_acc_GFC1}), and this mass is also used to update $M_{\rm cooled}$.


\subsubsection{\lgalaxy cooling model}
This cooling model is described in many \lgalaxy papers \citep[e.g.][]{munich_model1,munich_model2,munich_model_Guo11,munich_model3}. It assumes that at the start of a timestep, the hot gas is always distributed from $r=0$ to $r=r_{\rm vir}$, with a singular isothermal density profile, $\rho_{\rm hot}\propto r^{-2}$, and a single temperature, $T_{\rm vir}$. Both $r_{\rm vir}$ and $T_{\rm vir}$ are the values for the current halo. The total mass in the density profile is the current hot gas mass, $M_{\rm hot}$, and this fixes the normalization of this profile.

A cooling radius, $r_{\rm cool}$, is calculated using equation~(\ref{eq:r_cool_def}), with $t_{\rm cool,avail}=t_{\rm dyn}\equiv r_{\rm vir}/V_{\rm vir}$ (but note that early \lgalaxy papers, e.g.\ \citet{munich_model0}, made a different choice, and set $t_{\rm cool,avail}$ to be the age of the universe). If $r_{\rm cool}\leq r_{\rm vir}$, then the cooled down gas mass accreted onto the central galaxy within a timestep, $\Delta t$, is
\begin{eqnarray}
\Delta M_{\rm acc,gal} & = & 4\pi\rho_{\rm hot}(r_{\rm cool})\times r_{\rm cool}^2\frac{dr_{\rm cool}}{dt}\Delta t \nonumber \\
                       & = & \frac{M_{\rm hot}}{r_{\rm vir}}\frac{r_{\rm cool}}{t_{\rm dyn}}\Delta t,
\label{eq:M_gal_acc_lgalaxy}
\end{eqnarray}
with $dr_{\rm cool}/dt$ being estimated as $dr_{\rm cool}/dt=r_{\rm cool}/t_{\rm cool,avail}=r_{\rm cool}/t_{\rm dyn}$\footnote{We follow \citet{munich_model_Guo11} regarding factors of 2 in these equations, which differ from the versions in some earlier \lgalaxy papers}. If $r_{\rm cool}>r_{\rm vir}$, then
\begin{equation}
\Delta M_{\rm acc,gal}=\frac{M_{\rm hot}}{t_{\rm dyn}}\Delta t.
\end{equation}


\subsubsection{\MORGANA cooling model}
This cooling model is described in detail in \citet{morgana1} and \citet{morgana2}. The hot gas in a dark matter halo is assumed to be in hydrostatic equilibrium, and a cold gas halo similar to that in the new cooling model is also introduced. As in the new cooling model, in the limit of zero timestep length, the boundary separating the cooled down and the hot gas, which is also the inner boundary of the hot gas halo, is at radius $r_{\rm cool}$. The hot gas halo density and temperature profiles are determined by the assumption of hydrostatic equilibrium and that the hot gas between $r_{\rm cool}$ and $r_{\rm vir}$ follows a polytropic equation of state. This generally gives more complex profiles than those used in previously introduced cooling models, but typically the derived density profile is close to the cored $\beta$-distribution, $\rho_{\rm hot}(r)\propto 1/(r^2+r_{\rm core}^2)$, while the temperature profile is very flat and close to $T_{\rm vir}$. Therefore in this work, for simplicity we adopt the $\beta$-distribution and $T_{\rm vir}$ as the density profile and temperature of the hot gas halo for this model. The core radius, $r_{\rm core}$, is again calculated from the correlation described in Appendix~\ref{app:r_core_hot}.

In calculating the mass cooling rate $\dot{M}_{\rm cool}$, instead of cumulating the thermal energy lost by radiative cooling and estimating gas cooling histories as done in the new cooling model, the \MORGANA model assumes that at any time, each hot gas shell contributes to $\dot{M}_{\rm cool}$ according to its own cooling timescale, $t_{\rm cool}(r)$ [defined in equation~(\ref{eq:t_cool_def})]. Specifically, it is assumed that
\begin{equation}
\dot{M}_{\rm cool}=4\pi\int_{r_{\rm cool}}^{r_{\rm vir}}\frac{\rho_{\rm hot}(r)}{t_{\rm cool}(r)}r^2dr.
\label{eq:M_cool_dot_morgana}
\end{equation}
This equation is supplemented by another equation,
\begin{equation}
\dot{r}_{\rm cool}=\frac{\dot{M}_{\rm cool}}{4\pi\rho_{\rm hot}(r_{\rm cool})r_{\rm cool}^2}-c_{\rm s}(r_{\rm cool}),
\label{eq:r_cool_morgana}
\end{equation}
where $c_{\rm s}(r)$ is the sound speed. The first term in
equation~(\ref{eq:r_cool_morgana}) describes the increase of $r_{\rm
  cool}$ due to radiative cooling, based on the picture that all
cooled down gas within time interval $dt$ comes from a single shell,
so that $\dot{M}_{\rm cool} \, dt=4\pi\rho_{\rm hot}(r_{\rm
  cool})r_{\rm cool}^2dr_{\rm cool}$. This does not seem very
consistent with what is assumed in
equation~(\ref{eq:M_cool_dot_morgana}) for $\dot{M}_{\rm cool}$. The
second term in equation~(\ref{eq:r_cool_morgana}) describes the
contraction of the hot gas halo induced by the reduction of pressure
support due to cooling. Since the hot gas halo is in hydrostatic
equilibrium in the gravitational potential well of the dark matter
halo, $c_{\rm s}(r_{\rm cool})$ is comparable to the circular velocity
at $r_{\rm cool}$, so the contraction timescale is comparable to
$t_{\rm ff}(r_{\rm cool})$. Therefore, the contraction of the hot gas
halo in \MORGANA is similar to that assumed in the new cooling
model. Halo growth does not lead to any
  immediate adjustment of $r_{\rm cool}$ (unlike in the new cooling
  model), but does affect its subsequent evolution by changing the
  density and temperature profiles of the hot gas.

The mass of gas cooled down in one timestep is then $\Delta M_{\rm cool}=\dot{M}_{\rm cool}\Delta t$. It is used to update the mass of the cold gas halo, $M_{\rm halo,cold}$, and the mass accreted onto the central galaxy, $M_{\rm acc,gal}$, is then derived assuming gravitational infall of the halo cold gas, which is calculated in the same way as in the new cooling model, using equation~(\ref{eq:M_gal_acc_new_cool}).

The cooling model described in \citet{morgana1} includes additional suppression of cooling during halo major mergers, in which the cooling is forced to pause for several halo dynamical timescales. However, \citet{monaco_2014_comp} argued that this suppression of cooling seems to be too strong when compared with SPH simulations and suggested to turn it off. Here, for simplicity, we do not include this suppression in our implementation of the \MORGANA cooling model. We will discuss the effects of halo mergers on gas cooling and this suppression effect in \S\ref{sec:results_halo_mergers} and \S\ref{sec:results_model_comp_case_study} respectively.

\subsubsection{Cooling functions and halo mass threshold for cooling}
\label{sec:cooling_thresholds}
The radiative cooling functions $\tilde{\Lambda}(T)$ used for the five previously introduced SA cooling models are the same as those for the hydrodynamical simulations, which are described in \S\ref{sec:method_simulations}.

In the hydrodynamical simulations we impose a temperature floor for radiative cooling (\S\ref{sec:method_simulations}) to prevent significant gas cooling in halos less massive than $2\times 10^{10}\Msol$. Therefore we impose a corresponding halo mass threshold in the SA cooling models, and the gas in halos with $M_{\rm halo}<2\times 10^{10}\Msol$ is not allowed to cool.


\section{Results} 
\label{sec:results}

In \S\ref{sec:results_cooling_physics} we investigate
  several aspects of the physics of gas cooling by comparing the
  predictions of the new cooling model with the simulations. Then in
  \S\ref{sec:results_model_comp_case_study} and
  \S\ref{sec:stat_comparison} we compare predictions from the different
  SA cooling models described in \S\ref{sec:method_SA_models} with our
  hydrodynamical simulations. In
  \S\ref{sec:results_model_comp_case_study} we investigate some
  details of the SA models through case studies, because these details
  are not clearly seen in a statistical analysis, while in
  \S\ref{sec:stat_comparison} we perform a statistical comparison.

\subsection{Physics of gas cooling in halos} 
\label{sec:results_cooling_physics} 
SA models employ a very simple picture for gas cooling, while hydrodynamical simulations provide more complex detail. Comparing the predictions from these two methods can highlight some important aspects of the physics of gas cooling. In this subsection we compare simulation predictions with SA models for several individual halos. The simulation predictions are from the hydrodynamical simulation in the $25\,\Mpc$ cube described in \S\ref{sec:method_simulations}, while the SA model used here is the new gas cooling model introduced in \S\ref{sec:method_SA_models_new_cool}, which we have argued is the most physically realistic SA cooling model.

\subsubsection{Fast Cooling Regime vs. Filamentary Accretion} \label{sec:results_cold_vs_filamentary} 
\begin{figure*}
	\centering
	\includegraphics[width=0.7\textwidth]{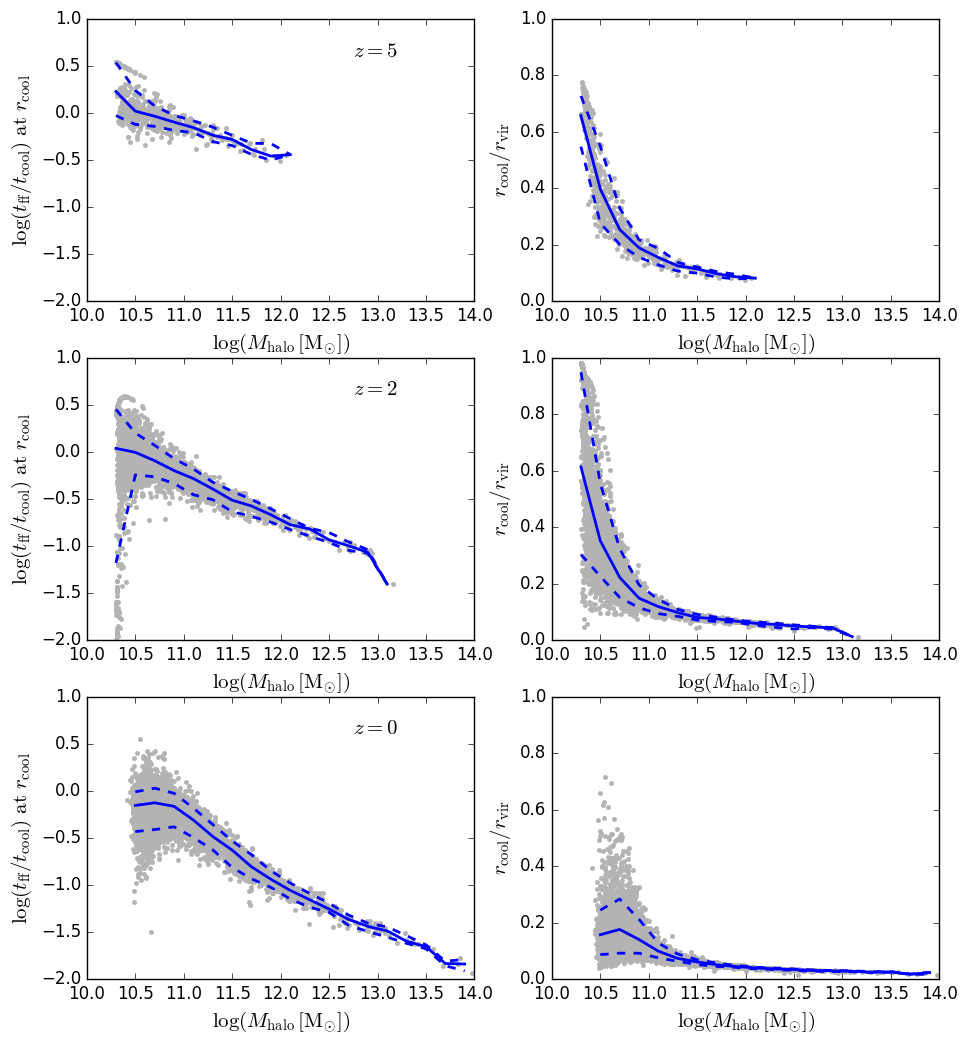}
	\caption{{\it Left panels:} the ratio of the free-fall timescale, $t_{\rm ff}$, to the cooling timescale, $t_{\rm cool}$, predicted by the new SA cooling model for individual halos from the simulation at redshifts $z=0,\ 2,\ 5$. Both timescales are calculated at the cooling radius, $r_{\rm cool}$, predicted by the same SA cooling model. {\it Right panels}: the ratio of the cooling radius, $r_{\rm cool}$, to the halo virial radius, $r_{\rm vir}$, predicted for the same halos. The halo sample here includes all halos from the simulation more massive than $3\times 10^{10}\Msol$ at $z=0$, and their most massive progenitors at $z=2$ and $z=5$. In all panels, gray dots are for individual halos, while the solid blue lines show the medians at a given halo mass, and the dashed blue lines indicate the $10-90$ percentiles.}

\label{fig:t_tt_2_t_cool_ratio} 
\end{figure*}

SA models generally predict that for low mass halos, the gas
radiatively cools faster than it falls in under gravity, and so a part
of the halo gas should be cold. This is called the fast cooling
regime.  This regime ends when the cooling timescale becomes
significantly longer than the free-fall timescale, and afterwards a
hot gas halo becomes dominant. This is called the slow cooling
regime. While this roughly defines the boundary between the two
regimes, the exact boundary is somewhat arbitrary, and the criterion
for it is likely to depend on the SA cooling model used. For our new
cooling model, we find it convenient to define halos as being in the
slow cooling regime for $t_{\rm cool}(r_{\rm cool})/t_{\rm ff}(r_{\rm
  cool}) > 3$, where $t_{\rm ff}(r)$ and $t_{\rm cool}(r)$ are
respectively the free-fall and cooling timescales at radius $r$, and
$r_{\rm cool}$ is the cooling radius calculated from the new cooling
model. The factor $3$ in this definition is somewhat arbitrary, but
our overall conclusions are not affected by modest variations in this
factor. In Fig.~\ref{fig:t_tt_2_t_cool_ratio} we plot the ratios
$t_{\rm ff}(r_{\rm cool})/t_{\rm cool}(r_{\rm cool})$ and $r_{\rm
  cool}/r_{\rm vir}$ predicted by the new cooling model for a sample
of halos from the simulation at redshifts $z=0,\ 2,\ {\rm and}\
5$. According to our criterion, halos with $t_{\rm ff}(r_{\rm
  cool})/t_{\rm cool}(r_{\rm cool}) < 1/3$ are defined to be in the slow cooling regime. From the figure, this criterion is seen to correspond to halo masses $M_{\rm halo}\gsim 2\times 10^{11} - 10^{12} \Msol$, with the threshold increasing slowly with redshift over the range $0<z<5$. We also see from the figure that for $M_{\rm halo}\gsim 2\times 10^{11} - 10^{12} \Msol$, the cooling radius, $r_{\rm cool}$, is always a small fraction of the halo virial radius, $r_{\rm vir}$, typically $r_{\rm cool}/r_{\rm vir} \lsim 0.1$. The low values of $r_{\rm cool}/r_{\rm vir}$ result in part from the gravitational contraction of the hot gas halo that forms part of the new SA cooling model.

Based on simulations, many previous works have argued for a more complex picture \citep[e.g.][]{cold_accretion_keres,dekel_filament_accretion}, in which the gas is delivered to dark matter halos through filaments rather than being spherically accreted, and in low mass halos, these cold filaments can reach all the way to the central galaxy, and so never build a spherical cold gas halo. Only at later times, when they become wider and less dense, and the halo has a higher virial temperature, do these filaments join the hot gas halo.

\begin{figure*}
	\centering
	\includegraphics[width=1.0\textwidth]{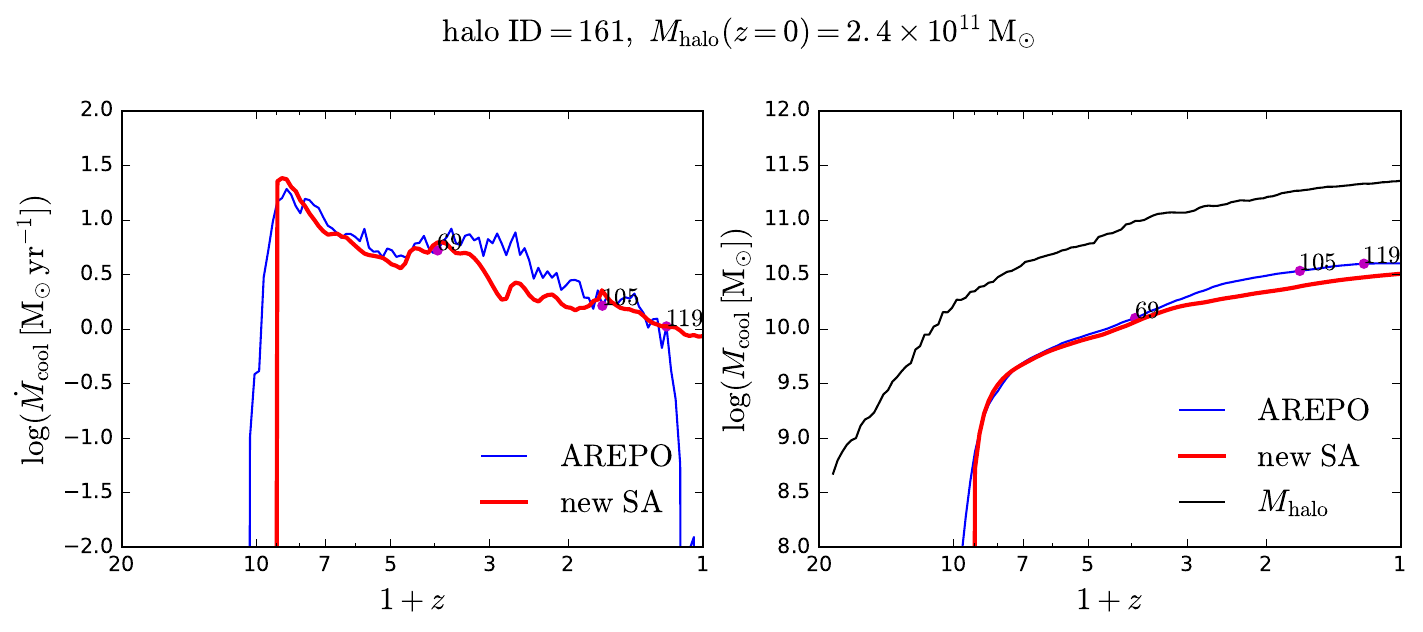}
	\caption{Comparison of mass cooling rates between a hydrodynamical simulation and the new SA cooling model for a single example halo of present-day mass $2.4 \times 10^{11} \Msol$ (and thus close to the upper limit for the fast cooling regime). {\it Left}: The predicted mass cooling rate. The sharp drop in the simulation result near $z=0$ is an artificial effect (for more details see \S\ref{sec:results_artificial_effects}). {\it Right}: The predicted cumulative cooled down mass. The growth of halo mass is also shown for reference in the right panel. In both panels, the magenta points label the snapshots for which the density and temperature distributions are shown in Fig.~\ref{fig:cooling_details_fast_cooling1}, and the associated numbers are the snapshot IDs.}

\label{fig:cooling_results_fast_cooling1} 
\end{figure*}

These two pictures are very different for low mass halos ($M_{\rm
  halo}\lsim 3\times 10^{11}\Msol$). How important this is for galaxy
properties depends on the cooled gas masses that they
predict. Fig.~\ref{fig:cooling_results_fast_cooling1} compares the
cooling history of a low mass halo (halo ID 1161) predicted by the new SA model and
by the hydrodynamical simulation. In the figure we
  compare both the mass cooling rate and the cumulative mass that has
  cooled. This halo has mass $2.4\times 10^{11}\Msol$ at $z=0$, and
so is close to the upper mass limit for the fast cooling regime only
at very late times. 

\begin{figure*}
	\centering
	\includegraphics[width=1.0\textwidth]{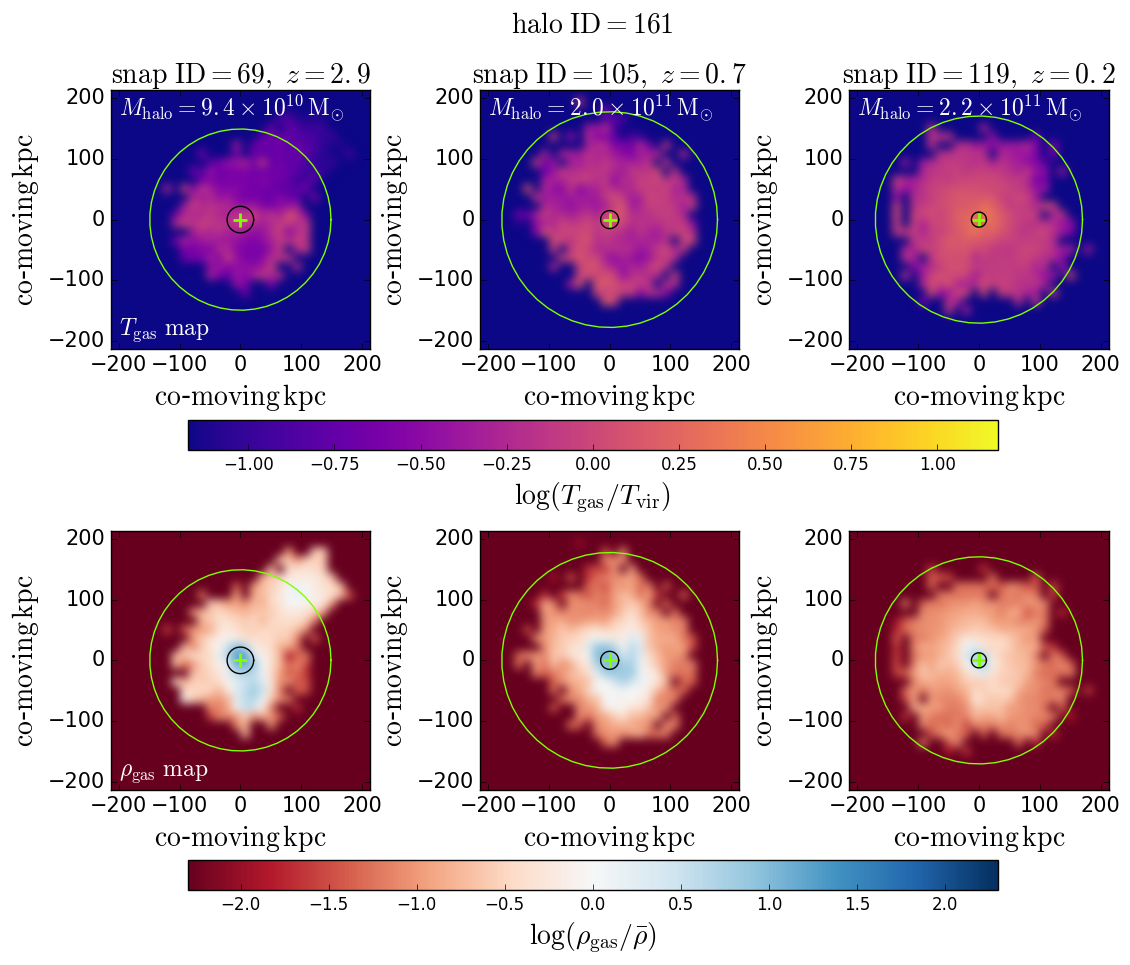}
	\caption{The projected gas temperature (top row) and density
          (bottom row) distributions of halo $161$ at selected
          snapshots. The snapshot IDs and redshifts are shown along
          the top, and the corresponding halo masses are given in the
          top row of panels. The density is expressed in units of the
          mean baryon density of the halo, $\bar{\rho}=(\Omega_{\rm
            b0}/\Omega_{\rm m0})\Delta'_{\rm vir}\rho_{\rm
            crit}$. Each pixel in a map shows the averaged temperature
          or density of the gas cells along the line of sight, with
          the average being weighted by the gas cell masses. In each
          panel, the green cross shows the halo centre, the green
          outer circle indicates $r_{\rm vir}$, and the black inner
          circle shows the cooling radius $r_{\rm cool}$ predicted by
          the new SA cooling model. These halos have
            $T_{\rm vir} \sim 2\times 10^5\,{\rm K}$, meaning the
            filamentary gas has temperature $\sim 10^4\,{\rm K}$.}

\label{fig:cooling_details_fast_cooling1} 
\end{figure*}

This figure shows that the predictions for the mass cooling rate from the new SA model and from the simulation are generally in good agreement for this halo. The gas cooling is seen to turn on suddenly at $z \sim 8-9$ in both the gas simulation and the SA model. This is a result of the temperature threshold set for gas cooling in the gas simulation, and the corresponding halo mass threshold set in the SA model, as described in \S\ref{sec:cooling_thresholds}. The large drop in mass cooling rates at $z\sim 0$ seen in the simulation results is, however, an artificial effect that results from our method of estimating the mass of gas that has cooled down over a timestep from the mass of stars formed in that timestep, as will be discussed later in \S\ref{sec:results_artificial_effects}. After allowing for this, the predicted cumulative cooled down masses for the SA model and the simulation are also generally in good agreement.

To see further details of the gas cooling for this halo, we select three snapshots and plot the projected gas temperature and density distributions. These selected snapshots are labeled as magenta dots in Fig.~\ref{fig:cooling_results_fast_cooling1}, and the corresponding gas distributions are shown in Fig.~\ref{fig:cooling_details_fast_cooling1}.

According to Fig.~\ref{fig:cooling_details_fast_cooling1}, at high
redshift, $z\sim 3$, the gas is clearly
  filamentary, as can be seen in
the density map, and the temperature map indicates that the filament gas
  is cold, with $T\sim 0.1T_{\rm vir}$. This confirms the findings in
previous works
\citep[e.g.][]{cold_accretion_keres,dekel_filament_accretion}. Only at
later times, when $z\sim 0.7$, does the gas distribution become more
spherical, and closer to the picture in the SA model. At $z=0.7$,
there is still an obvious halo cold gas component. The gas
distribution becomes more spherical because at low redshift, the
filaments become very wide, with radius comparable to $r_{\rm vir}$ of
the halo, and so the accretion is close to spherical
\citep[e.g.][]{cold_accretion_keres}. Then even later, at $z=0.2$, the
hot gas halo begins to appear, which indicates the transition from a
cold halo gas to a hot gas halo. This transition happens at a mass
around $2.2\times 10^{11}\Msol$ for this halo, which is close to the
SA model prediction, which is around $2\times 10^{11}\Msol$.

Although the simulation gives gas distributions very different from the SA model for $z>0.7$, the predicted mass cooling rates are similar. This is because for this case, in both the simulation and the SA model, the gas accretion onto the central galaxy occurs on a timescale close to the free-fall timescale. In the simulation, the gas is delivered by the cold filaments, which are difficult to heat up, and is expected to fall freely onto the central galaxy under gravity. On the other hand, in the SA model, although it is assumed that the gas accreted onto the dark matter halo is nominally shock heated to build a hot gas halo, in the fast cooling regime the cooling is very efficient, and the accretion onto the central galaxy is either limited by the gravitational infall timescale or set by a cooling timescale that is comparable to the former, so the timescale of this accretion is always close to the gravitational infall timescale.

\begin{figure*}
	\centering
	\includegraphics[width=1.0\textwidth]{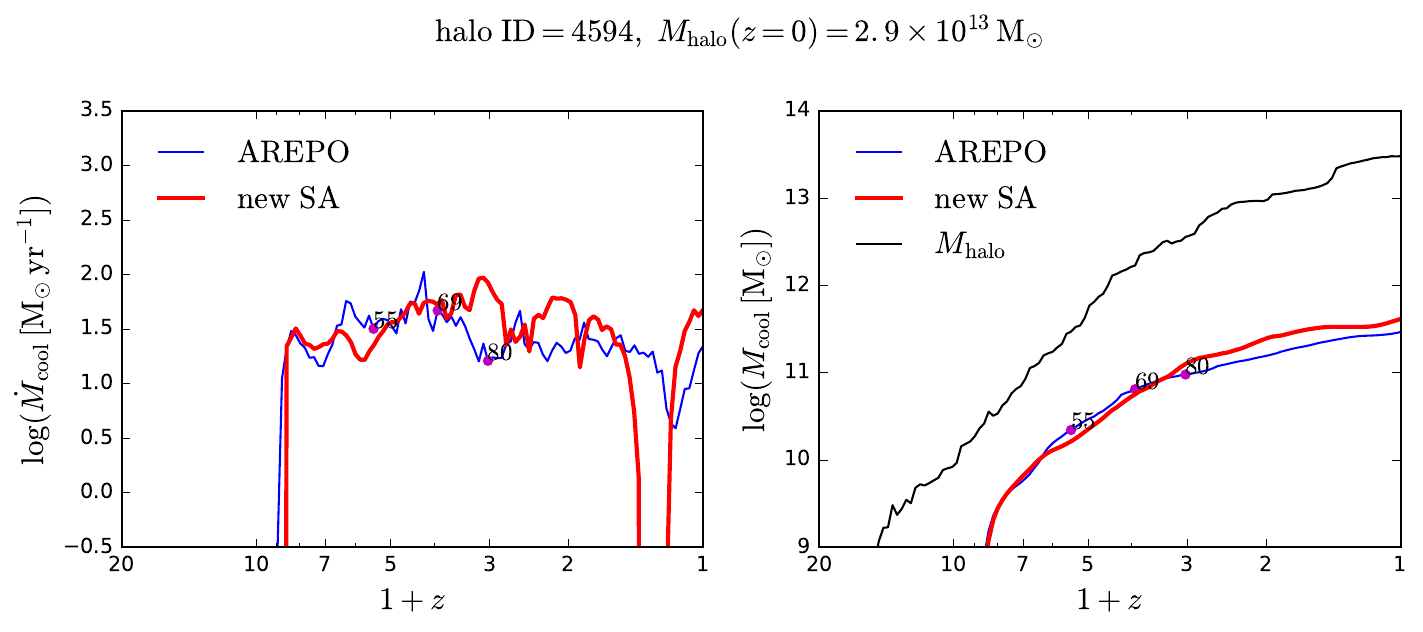}
	\caption{The cooling histories predicted by the SA model and simulation for halo with ID $4594$. The meaning of the labels is the same as in Fig.~\ref{fig:cooling_results_fast_cooling1}, and for more information see the caption there. The magenta points indicate the snapshots for which temperature and density distributions are plotted in Fig.~\ref{fig:cooling_details_fast_cooling2}.}

\label{fig:cooling_results_fast_cooling2} 
\end{figure*}

Low mass halos at high redshift can also be the progenitors of massive halos at low redshift. Compared to the case studied above, in which the halo remains low in mass down to $z\sim 0$, these progenitors are formed in very different environments, so the gas accretion can be different. Fig.~\ref{fig:cooling_results_fast_cooling2} shows the cooling history of a massive halo (halo ID 4594), with mass $2.9\times 10^{13}\Msol$ at $z=0$. Here we will focus on the relatively high redshift range ($z\gsim 2$), for which the halo mass is still low. The figure shows that at $z>2$ the predictions from the SA model and the simulation are in good agreement, although this agreement of the predicted mass cooling rates is not as good as that for the lower mass halo studied above.

\begin{figure*}
	\centering
	\includegraphics[width=1.0\textwidth]{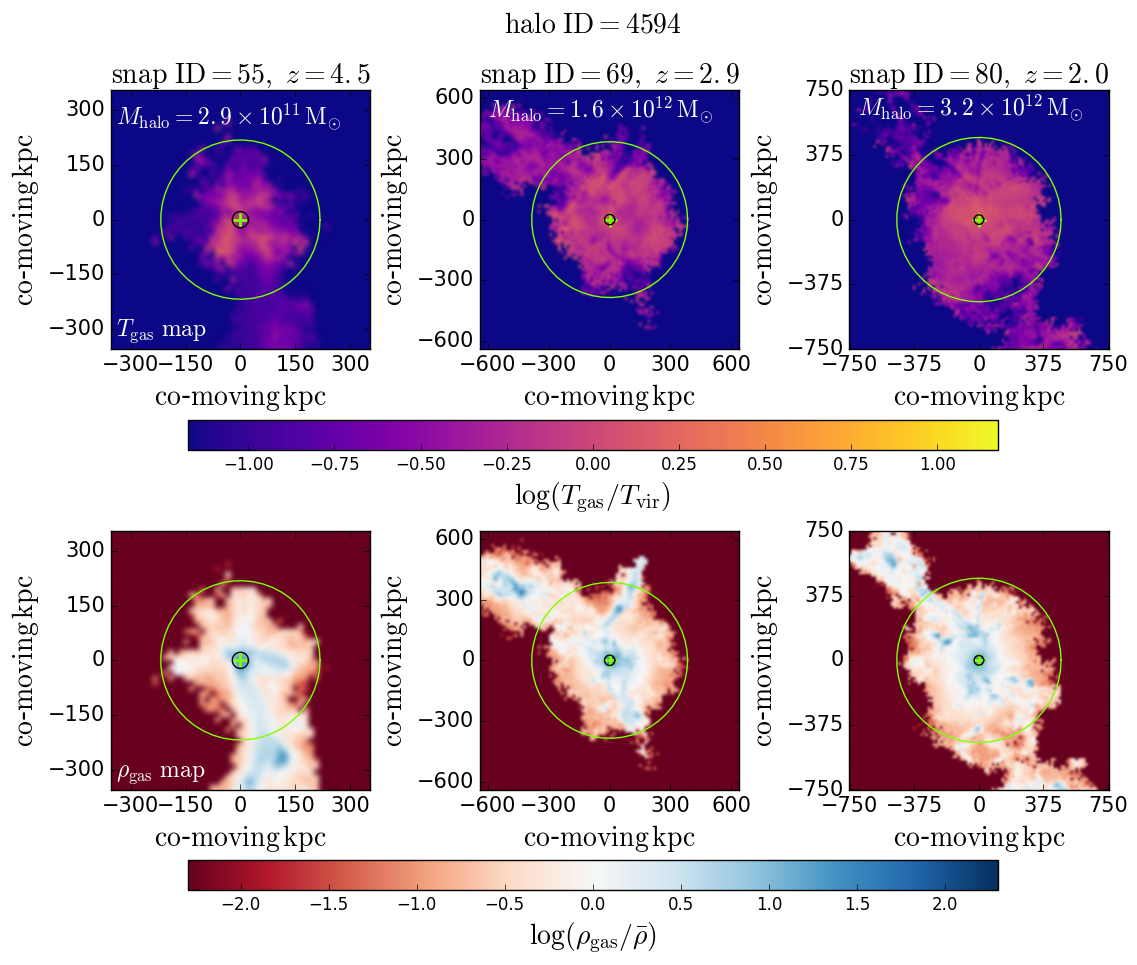}
	\caption{The projected gas temperature and density
          distributions of halo $4594$ at selected snapshots indicated
          in Fig.~\ref{fig:cooling_results_fast_cooling2}. The top row
          shows temperature maps while the bottom row shows density
          maps. The meaning of the labels and colour scales is the
          same as in Fig.~\ref{fig:cooling_details_fast_cooling1}, and
          for more information see the caption to that figure.
          These halos have $T_{\rm vir} \sim
            10^6\,{\rm K}$, meaning the filamentary gas has
            temperature $\sim 10^5\,{\rm K}$. }

\label{fig:cooling_details_fast_cooling2} 
\end{figure*}

At $z>4$, $M_{\rm halo}\lsim 3\times 10^{11}\Msol$, and so according to the criterion discussed in \S\ref{sec:results_cold_vs_filamentary},  the SA model predicts the halo to be in the fast cooling regime. Between $z=4$ and $z=2$, the halo grows in mass from $3\times 10^{11}\Msol$ to about $3\times 10^{12}\Msol$, which is roughly in the transition range from fast cooling to slow cooling. To further investigate the details of cooling for this halo at $z>2$, we select three snapshots and show the corresponding gas distributions in Fig.~\ref{fig:cooling_details_fast_cooling2}. At $z=4.5$, the gas is obviously filamentary. It is also cold, because the temperature map indicates that it has $T_{\rm gas}\lesssim 0.1T_{\rm vir}$.  Later on, this  gas halo  gradually evolves to a more spherical shape. At $z=2.9$, a hot halo has appeared, but its temperature seems to be slightly lower than $T_{\rm vir}$ (purple patches appear within $r<0.5r_{\rm vir}$, and we have checked that this is not due to projection effects). By $z=2$, the hot gas has become hotter, with temperature closer to $T_{\rm vir}$. This transition is different from the simple picture in the SA model, because of the non-spherical filaments, but the SA model still manages to predict roughly the correct cooling history, at least for this specific case. In particular the final cooled down mass in the SA model is very close to that in the hydrodynamical simulation.

In summary, filamentary accretion is commonly seen at high redshifts
($z\gsim 2$), but insofar as the timescale of the accretion onto the
central galaxy is comparable to the free-fall timescale, the simple
spherical gas cooling picture in the SA model does not much degrade
the predictions for the mass cooling rates. It seems that the SA model
also gives roughly the correct cooling histories during the transition
from anisotropic filaments to a spherical hot gas halo, at least for
the individual halos that we have studied here. There
  appears to be a rough correspondence between the regimes of halo
  mass and redshift in which the halo gas in the simulation is
  dominated by cool filaments, and the regime in the SA model in which
  the halo is in the fast cooling regime, defined for our model as
  when $t_{\rm ff}(r_{\rm cool})/t_{\rm cool}(r_{\rm cool}) > 1/3$. As
  already mentioned, for our SA cooling model, this criterion
  corresponds to a fairly well defined halo mass, increasing from $M_{\rm halo}\sim
  2\times 10^{11} \Msol$ at $z=0$ to $M_{\rm halo}\sim
  10^{12} \Msol$ at $z=5$, below which halos are in the fast
  cooling regime. Visual inspection of images like those shown in
  Fig.~\ref{fig:cooling_details_fast_cooling1} and
  Fig.~\ref{fig:cooling_details_fast_cooling2} suggest that gas in halos in
  the simulations transitions from filament-dominated to
  hot-halo-dominated at similar halo masses, although the correspondence is not exact, with the transition mass in the simulations appearing to be somewhat larger than in the SA model at higher redshifts.

A comparison of the transition between fast (cold)
  and slow (hot) accretion in simulations and SA models was previously
  made by \citet{Lu_2011_comp}. \citeauthor{Lu_2011_comp} considered a
  number of different SA models, each with their own criterion for
  separating fast and slow accretion, and an SPH simulation, for which
  gas accretion onto the central galaxy was separated into cold and
  hot components, and compared average accretion rates as a function
  of halo mass and redshift. They found that for most of the SA
  models, the transition from fast to slow accretion happened at a
  lower halo mass than the transition from cold to hot accretion in
  the simulation. However, the condition in our SA model typically
  places the transition from fast to slow cooling at a somewhat larger
  value of $t_{\rm cool}(r_{\rm cool})/t_{\rm ff}(r_{\rm cool})$ than
  for most of the SA models considered by \citeauthor{Lu_2011_comp},
  corresponding to larger halo masses, and so should be more
  consistent with their SPH simulation results, if one identifies fast
with cold accretion, and slow with hot accretion.


\subsubsection{Slow cooling Regime} 
\label{sec:results_slow_cooling_regime} 
When a halo is massive enough, the SA model predicts the cooling timescale to be much longer than the dynamical timescale, and the hot gas in the dark matter halo forms a quasi-hydrostatic hot gaseous halo, from which the gas cools and falls onto the central galaxy. This is the so-called slow cooling regime. Typically the hot gas halo has higher density at smaller radii, so the inner part of the halo cools more rapidly. Therefore, a naive expectation is that the temperature of the gas should decrease with radius, with the outer part having temperature close to $T_{\rm vir}$ as it experiences less cooling, while the inner part contains partly cooled gas, and the cold gas (fully cooled down) is in the halo centre.

\begin{figure*}
	\centering
	\includegraphics[width=1.0\textwidth]{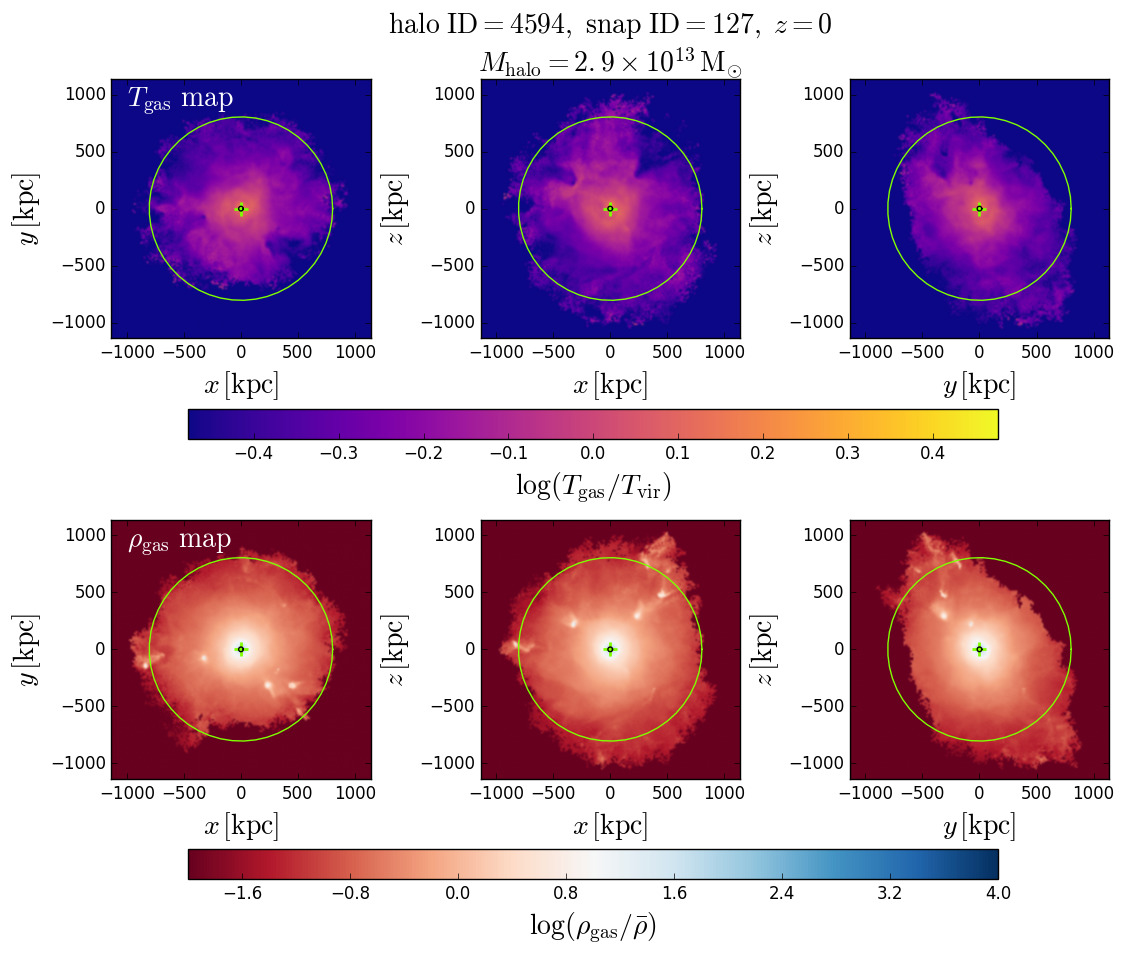}
	\caption{The projected halo gas temperature (upper row) and
          density (lower row) maps for halo $4594$ at $z=0$. Each
          pixel in the map show the averaged temperature or density of
          the gas cells along the line of sight. The average is
          weighted by the gas cell masses. The top row shows
          temperature maps, projected onto the $xy$, $xz$ and $yz$
          planes, while the bottom row shows the corresponding density
          maps. The gas density is in units of $\bar{\rho}$, where
          $\bar{\rho}=(\Omega_{\rm b0}/\Omega_{\rm m0})\Delta'_{\rm
            vir}\rho_{\rm crit}$ is the mean baryon density of the
          halo. These maps show a roughly spherical gaseous halo with
          temperature around $T_{\rm vir}=5.5\times
            10^6  \,{\rm K}$. In each panel, the green
          cross shows the halo centre, the green outer circle
          indicates $r_{\rm vir}$, and the black inner circle shows
          the cooling radius, $r_{\rm cool}$, predicted by the new SA
          cooling model. }

\label{fig:slow_cooling_maps}
\end{figure*}

To compare the above picture with hydrodynamical simulations, in Fig.~\ref{fig:slow_cooling_maps} we plot the gas density and temperature distributions at $z=0$ for the same halo whose cooling history is shown in Fig.~\ref{fig:cooling_results_fast_cooling2}. At $z=0$, this halo has mass about $3\times 10^{13}\Msol$, which is massive enough to be in the slow cooling regime predicted by the SA model. Fig.~\ref{fig:slow_cooling_maps} shows the projected halo gas temperature and density maps of this halo at $z=0$. It is clear from this figure that there is a more or less spherical gas halo with temperature close to $T_{\rm vir}$. Thus, the qualitative expectation from the SA cooling model is confirmed.

\begin{figure*}
	\centering
	\includegraphics[width=1.0\textwidth]{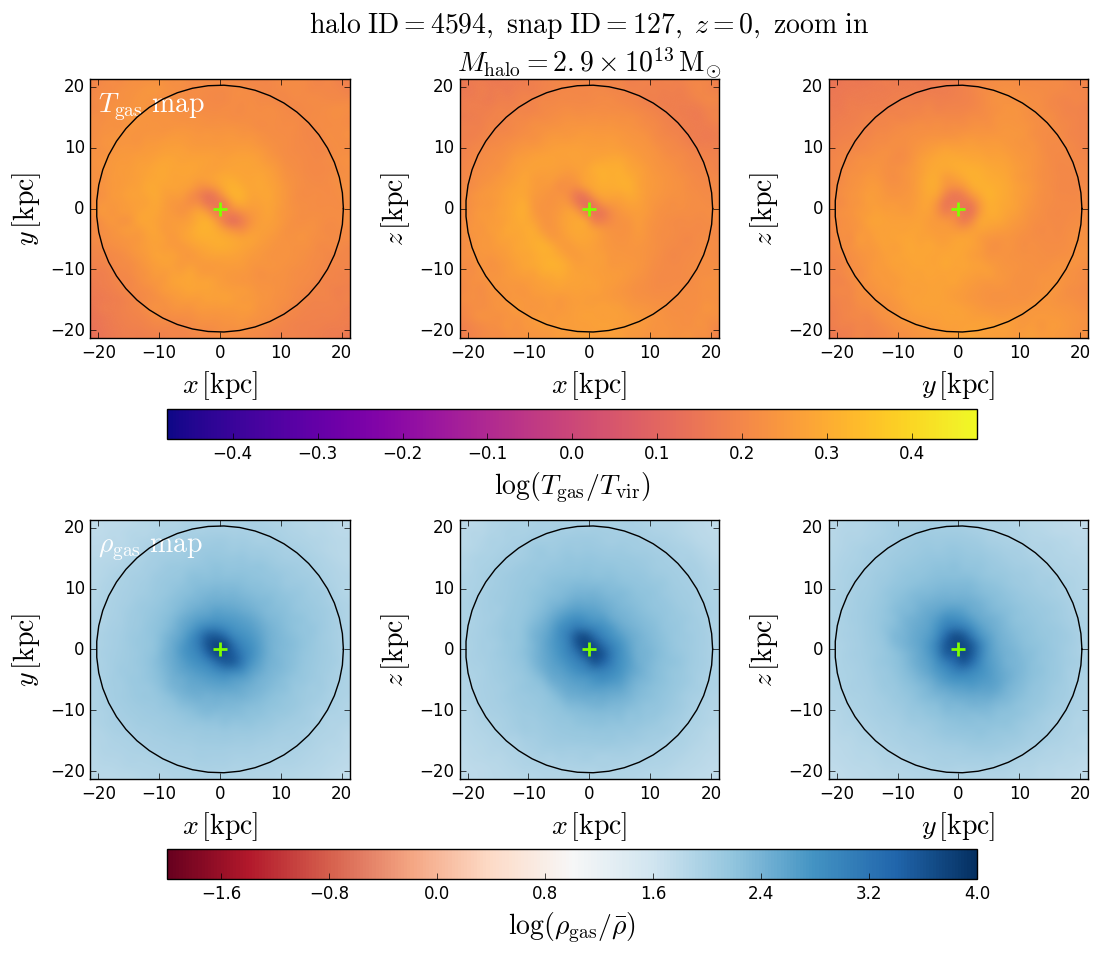}
	\caption{The projected halo gas temperature (upper row) and density (lower row) maps for the central region of the halo shown in Fig.~\ref{fig:slow_cooling_maps}, at the same redshift, $z=0$. The meanings of maps and symbols are the same as in Fig.~\ref{fig:slow_cooling_maps}, and for more information see the caption to that figure. The black circles are the same as the inner black circles plotted in Fig.~\ref{fig:slow_cooling_maps}, and indicate the cooling radius predicted by the new SA cooling model.}

\label{fig:slow_cooling_maps_zoom_in}
\end{figure*}

However, the temperature maps do not show a temperature decreasing with radius, but instead, the temperature is always close to $T_{\rm vir}$, and is even higher at smaller radii. To further investigate the details in the central region of the halo, we generated maps of projected density and temperature for the central region, which are shown in Fig.~\ref{fig:slow_cooling_maps_zoom_in}. This figure further confirms that there is no gas with temperature significantly lower than $T_{\rm vir}$ in the central region of the gaseous halo.
It also shows that in the very central region, the gas becomes very dense while keeping its temperature close to $T_{\rm vir}$, and a disky gas structure forms, with  density $10^4$ times higher than the mean baryon density of the whole halo.

Since there are newly formed stars in this halo at this timestep, there must be gas cooling, but the temperature maps suggest that the gas keeps a roughly constant temperature during cooling. This means there must be heating sources to balance the radiative cooling. Since in the current simulation there are no feedback processes, the only possible heating source is the gravitational potential energy. When the gas in the halo centre finally cools down, it no longer provides pressure support to the hot gas halo, and this causes the latter to contract towards the halo centre. During this contraction, gravity does positive work on every shell by compressing the gas, and this balances the energy losses due to cooling. This process continues until the gas reaches the very central region, where the radius is small enough that the gas becomes centrifugally supported due to its angular momentum, which  halts further infall. At this stage, the gas has reached very high density (as indicated by Fig.~\ref{fig:slow_cooling_maps_zoom_in}), and radiates its thermal energy on a very short timescale, and so becomes cold gas. This picture for the gas cooling was previously mentioned in \citet{morgana2} [see also \citet{compression_heating}].

According to this picture, when a gas shell moves from the outer region to the halo centre, the cooling effectively radiates away the contraction work done by gravity. Since the temperature maps show that the gas has a roughly constant temperature around $T_{\rm vir}$, this contraction can be  treated as roughly isothermal. Then the total compression work done on the gas for a shell with original radius $r$ is
\begin{eqnarray}
W(r) & = & -\int_{V(r)}^{V(0)} PdV = m_{\rm gas}\int_{\rho(r)}^{\rho(0)} \frac{P}{\rho^2} d\rho \nonumber \\
& = & \frac{k_{\rm B}T_{\rm vir}m_{\rm gas}}{\mu_{\rm m}} \int_{\rho(r)}^{\rho(0)} \frac{1}{\rho}d\rho \nonumber \\
& = & \frac{2}{3}U\ln \frac{\rho(0)}{\rho(r)}, \label{eq:compression_work}
\end{eqnarray}
where $V$ is the volume of this gas shell, $m_{\rm gas}$ its mass and $\rho$ its density, $k_{\rm B}$ is the Boltzmann constant, $\mu_{\rm m}$ the mean molecular mass, and $U=(3k_{\rm B}T_{\rm vir}m_{\rm gas})/(2\mu_{\rm m})$ is the thermal energy of the shell. With the shell density given by the $\beta$-distribution with $r_{\rm core}\sim 0.1r_{\rm vir}$, and for the most extreme case in which $r=r_{\rm vir}$, the above equation gives an upper limit $W\sim 3U$. Since $W(r)$ depends on $r$ through $\ln \rho(r)$, it depends only weakly on the starting radius, $r$, and thus the upper limit provides a rough estimate of the typical value of $W(r)$. At the halo centre, the gas radiates away its thermal energy $U$ and cools down. As described in the previous paragraph, this cooling in the halo centre is very fast, so here we approximate it as being instantaneous. Therefore, the time that it takes for a hot gas shell to cool down is approximately the time needed for this shell to radiate away the total compression work, $W$, done on it.

Therefore, instead of the simple picture assumed in most of the SA models, in which the gas radiates away its thermal energy $U$ and then cools down, the slow gravitational contraction instead requires the gas to have enough time to radiate away  $\sim 3U$ to cool down. The SA model therefore tends to overestimate the cooling rate in this regime. As will be discussed later in \S\ref{sec:stat_comparison} (see Fig.~\ref{fig:SA_stat_comp}), for halos with $M_{\rm halo}(z=0)>3\times 10^{11}\Msol$ and at $z\lesssim 1$, the mass cooling rates predicted by the new cooling model and by the \lgalaxy and \MORGANA models are a factor $\sim 2$ larger than those given by the gas simulation. Our analysis here provides a possible explanation for this effect.

Note that heating by gravitational contraction does not play an important role in the fast cooling regime, because there the cooling timescale is shorter than or comparable to the dynamical timescale, and so the gas completely cools down before significant contraction can happen, the pressure $P$ drops to very low values, and little $PdV$ work is done as the cold gas falls to the centre of the halo. Instead, the  gravitational  potential energy released is converted into kinetic energy of the infalling cold gas. So for the fast cooling regime, the gas still only needs to radiate away total energy $U$ in order to cool down. The gravitational contraction work is also not important if the gas is delivered by cold filaments.


\subsubsection{Effects of Halo Mergers} \label{sec:results_halo_mergers}
Almost all SA gas cooling models assume that the gas newly accreted onto a dark matter halo is shock-heated to the virial temperature, $T_{\rm vir}$, of this halo. If a large amount of gas is accreted in a relatively short time, then this could cause significant heating of the hot gas already in the halo, and thus interrupt its cooling. Halo major mergers are the most common cause of this rapid gas accretion.

Different SA cooling models treat the effect of this newly accreted gas differently. The \MORGANA model explicitly quenches cooling for some time \citep[e.g.][]{morgana1}, while other models use more implicit modelling. We defer the comparison between SA models to \S\ref{sec:results_model_comp_case_study}, and here we focus on the comparison between the new SA cooling model and the hydrodynamical simulation.

\begin{figure*}
	\centering
	\begin{minipage}{0.5\textwidth}
		\includegraphics[width=0.9\linewidth]{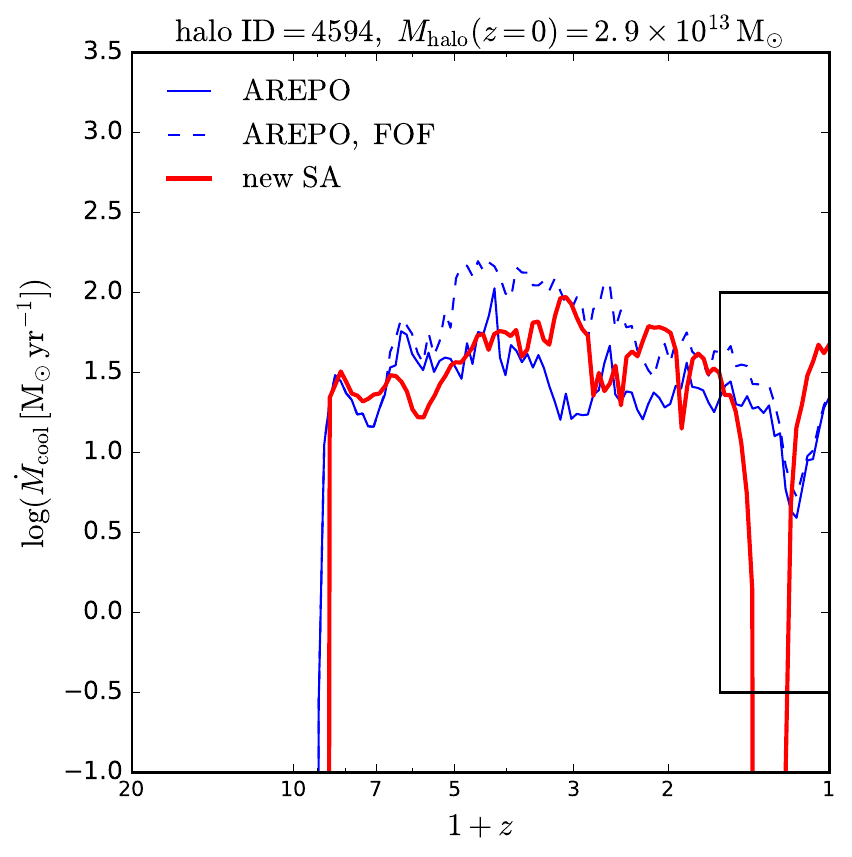}
	\end{minipage}%
	\begin{minipage}{0.5\textwidth}
		\includegraphics[width=0.9\linewidth]{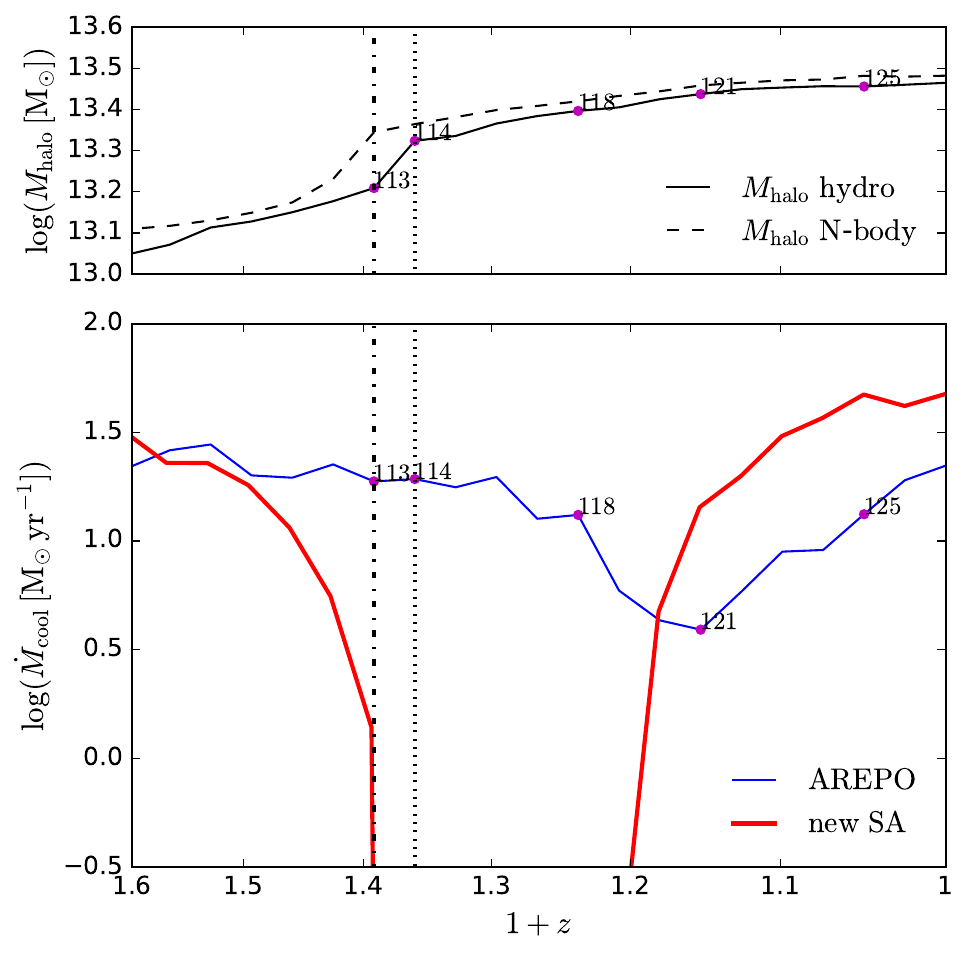}
	\end{minipage}
	\caption{{\it Left panel}: The mass cooling rate for halo $4594$. The blue lines show cooling rates measured from the hydrodynamical simulation, with the solid line being the measurement for the central galaxy only, while the dashed line is that for the whole FOF group. The red solid line is the result from the new SA cooling model. {\it Right panels}: Zoom-in plot of the region in the black box in the left panel. The lower right panel shows the mass cooling rates, with the line types as in the left panel, and the magenta dots labelling the selected snapshots for which further details are shown in Fig.~\ref{fig:merger_halo4594_details}, with the corresponding numbers labelling the snapshot IDs. The upper right panel shows the growth of halo mass. The solid and dashed lines are respectively the Dhalo masses measured from  the hydrodynamical and N-body simulations (see \S\ref{sec:merger_tree_construction} for more details). The vertical dotted line indicates the completion time of the halo merger in the hydrodynamical simulation according to the Dhalo merger tree, while the vertical dashed-dotted line indicates the corresponding time in the dark matter only simulation.} \label{fig:merger_halo4594}
\end{figure*}

\begin{figure*}
	\centering
	\includegraphics[width=0.65\textwidth]{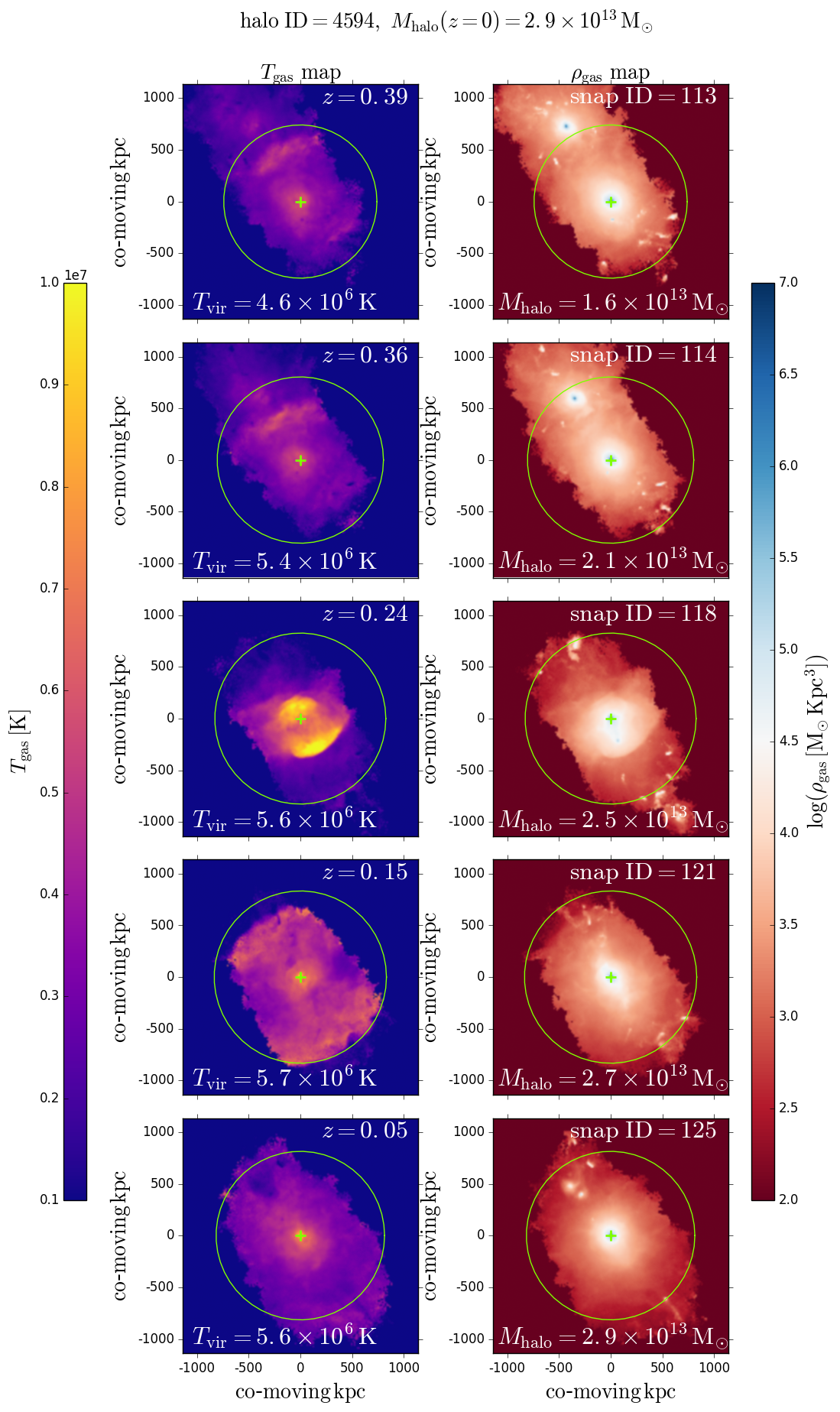}
	\caption{The projected temperature (left) and density (right) maps of the FoF group containing halo $4594$ for the selected snapshots labeled in Fig.~\ref{fig:merger_halo4594}, with time increasing downwards. To show the evolution more clearly, the colour scales for the temperature and density are absolute rather than scaled to halo properties. The colour scale for temperature is linear, in units of $K$, while the scale for density is logarithmic, with density in units of $\Msol\,{\rm kpc}^{-3}$. In each panel, the green circle indicates $r_{\rm vir}$ of the Dhalo, while the green cross shows its centre. The snapshot ID, redshift, halo mass and $T_{\rm vir}$ for each snapshot are given in each panel. } \label{fig:merger_halo4594_details}
\end{figure*}

In the new SA model, the effect of this newly accreted gas is modelled straightforwardly. This gas is assumed to be newly heated up to $T_{\rm vir}$ and thus has no previous cooling history. The accretion of it increases the total thermal energy of the hot gas halo, but leaves the total energy radiated away, $E_{\rm cool}$, unchanged. As a result, the cooling after the accretion could be suppressed, while the extent of this suppression depends on the amount of gas accreted. 

It is interesting to see whether this suppression of cooling expected in the new SA cooling model also occurs in the hydrodynamical simulation. Fig.~\ref{fig:merger_halo4594} shows the mass cooling rate as a function of redshift for halo $4594$, which ends up having a mass of $2.9\times 10^{13}\Msol$ at $z=0$. At $z\sim 0.4$, this halo experiences a major merger with mass ratio about $3:1$. Accordingly, the SA model prediction shows a sharp drop in the mass cooling rate. From the zoom-in plots on the right, it is clear that the sharp drop  in the SA model happens immediately after the merger. This is expected, because when using the Dhalo merger tree, the halo merger is treated as an instantaneous event, and in the SA model, the associated gas accretion and heating are also assumed to be instantaneous. In the simulation , there is also a drop, whereby the cooling rate is reduced by a factor about $5$. This drop is not as strong as that predicted by the SA model, but more importantly, it appears about $2$ halo dynamical timescales ($8$ snapshots) later than the merger. Although there are small differences between the halo growth histories in the dark matter only simulation (used in constructing merger trees for the SA model) and in the hydrodynamical simulation, as can be seen from the upper right panel of Fig.~\ref{fig:merger_halo4594}, this time delay is much larger than that, so it must be caused by something else.

To investigate further the details of the drop in the mass cooling rate, we extract the projected gas temperature and density maps for several snapshots covering the halo merger and the drop. The selected snapshots are labeled by magenta dots in the right panels of Fig.~\ref{fig:merger_halo4594}, and the maps are shown in Fig.~\ref{fig:merger_halo4594_details}. To better show the temperature evolution of the gas, a linear colour scale in absolute temperature is adopted. For each snapshot, the maps are for the whole FOF group that contains halo $4594$, to better show the pair of merging halos.

In snapshot $113$ ($z=0.39$), two merging halos are clearly seen in Fig.~\ref{fig:merger_halo4594_details}. They are in one FOF group, but the Dhalo algorithm still identifies them as two different Dhalos. In the temperature map, a region of weak heating due to gas compression can be seen between the two halos. Then in snapshot $114$ ($z=0.36$), these two halos become closer and form a single Dhalo, so in the merger tree this snapshot corresponds to the completion of the halo merger, but it seems that the merged structure has not yet relaxed, and the temperature map still only indicates a region of weak heating between these two merging halos. The merging process continues, and about one halo dynamical timescale ($4$ snapshots) later, in snapshot $118$ ($z=0.24$), a strong shock is generated by the merger and, from this moment on, the mass cooling rate begins to drop, as can be seen from the lower right panel of Fig.~\ref{fig:merger_halo4594}.

About one halo dynamical timescale later, at snapshot $121$ ($z=0.15$), the strong shock has expanded and heated up nearly the whole hot gas halo. Accordingly, the mass cooling rate drops to a minimum. From the density map, the gaseous halo appears to be largely relaxed by this time. Then, after about another halo dynamical timescale, at snapshot $125$ ($z=0.05$), the hot gas halo becomes cooler, and the mass cooling rate rises back to a level close to that before the merger.

Based on these maps, two points can be made. Firstly, the suppression
of cooling is associated with the shock heating induced by the merger,
so at least in this case, the halo major merger does suppress
cooling. Secondly, the suppression appears a few halo dynamical
timescales later after the completion of the Dhalo merger, because the
merging halos are identified as one Dhalo before they are fully
relaxed, and the time delay from the Dhalo merger to the suppression
period is roughly the halo dynamical timescale. The SA model 
assumes that the halo is relaxed as soon as the Dhalo
  merger has completed, thus shifting the drop in mass cooling rate to
  an earlier time than in the simulation, as well as predicting a
  stronger drop in the cooling rate than seen the simulation.
However, the general conclusion is that the SA model and
hydrodynamical simulation have similar behavior during the halo major
merger, and this is a success for the simple SA model.

\begin{figure*}
	\centering
	\includegraphics[width=0.6\textwidth]{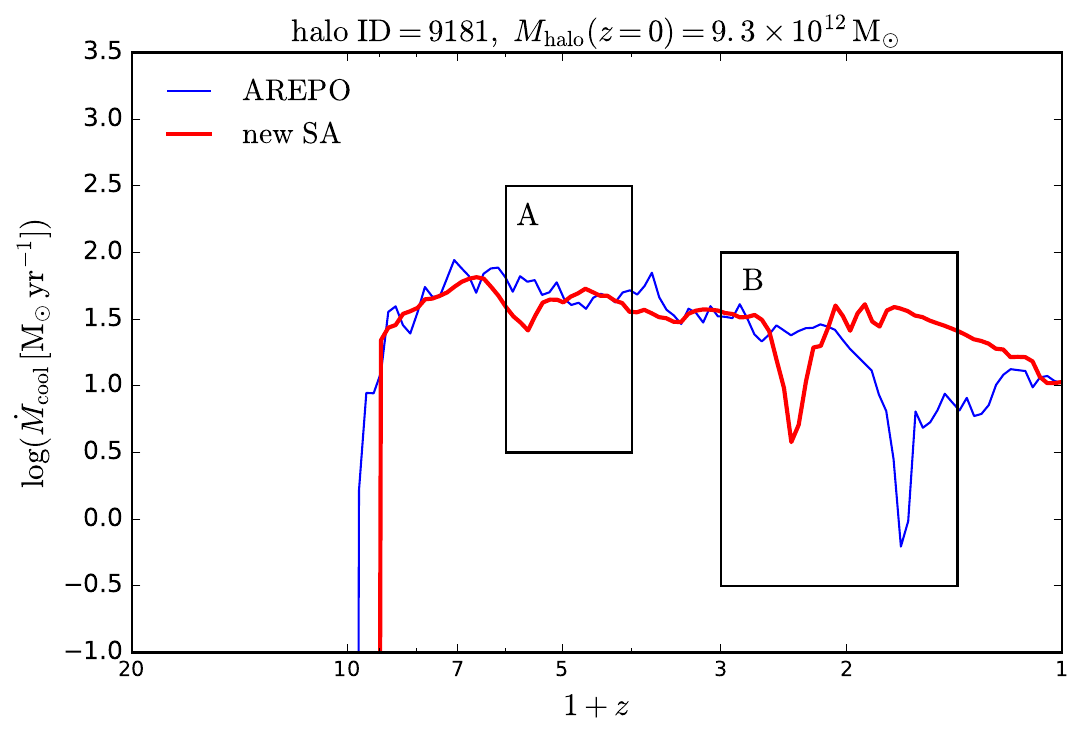}
	\begin{minipage}{0.5\textwidth}
		\includegraphics[width=0.9\linewidth]{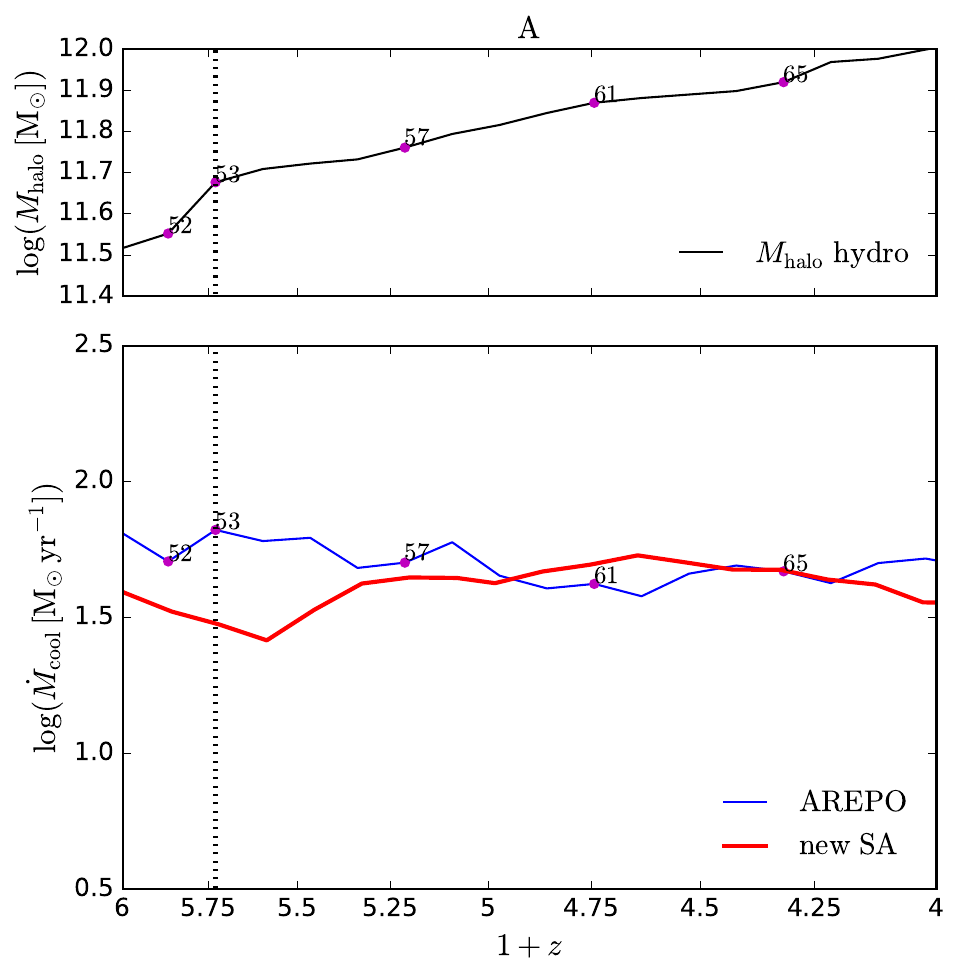}
	\end{minipage}%
	\begin{minipage}{0.5\textwidth}
		\includegraphics[width=0.9\linewidth]{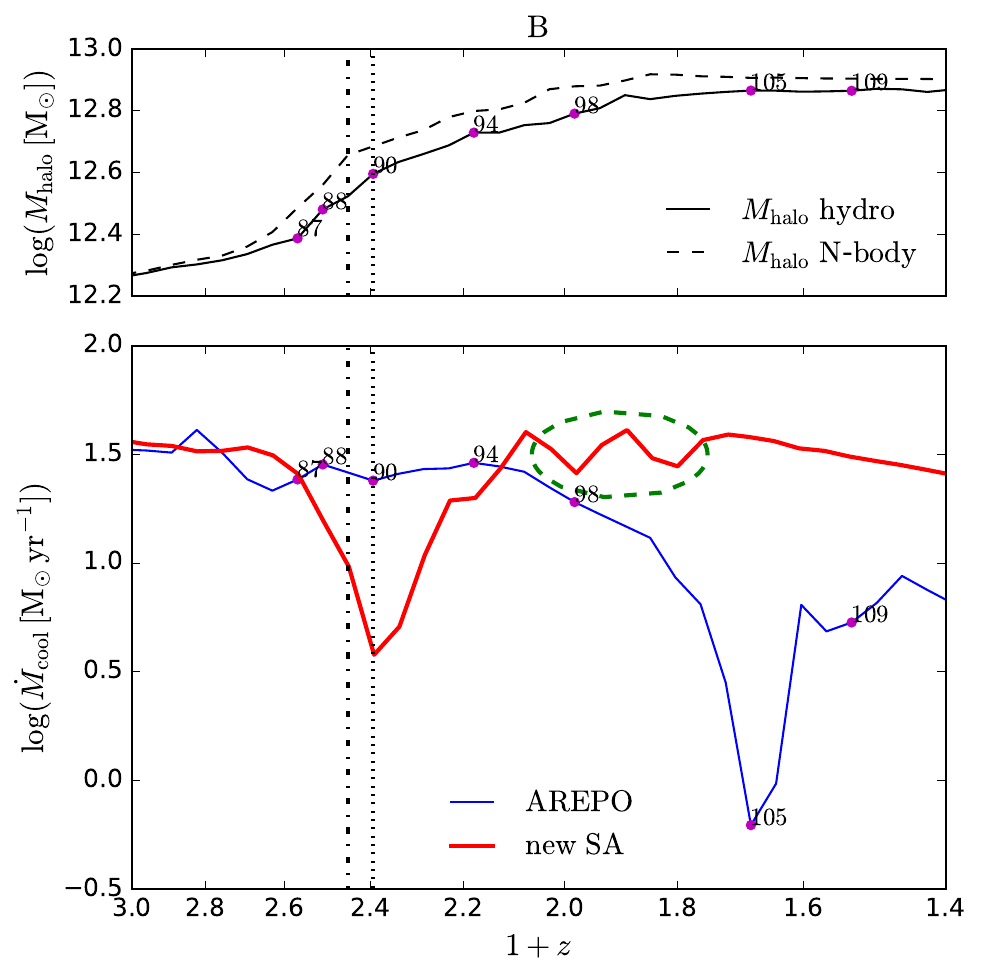}
	\end{minipage}
	\caption{{\it Top panel}: The mass cooling rate for halo $9181$. The blue solid line is the cooling rate measured from the hydrodynamical simulation, for the central galaxy only, while the red solid line is the prediction of the new SA cooling model. {\it Lower left panels}: Zoom-in plot of the region labeled `A' in the top panel. The magenta dots label the selected snapshots for which further details are shown in Fig.~\ref{fig:merger_halo9181_details_A}, and the associated numbers are the snapshot IDs. The small upper panel shows the growth in halo mass. In this case, the halo mass growth in the hydrodynamical and dark matter only simulations are almost the same, so only the result from the former simulation is plotted. {\it Lower right panels}: Zoom-in plot of the region labeled `B' in the top panel. The meanings of the symbols and lines are the same as for the lower left panels, but the magenta dots are for the snapshots plotted  in Fig.~\ref{fig:merger_halo9181_details_B}. In all of the lower panels, the vertical lines indicate the completion of the halo merger according to the Dhalo merger trees. The dotted and dashed-dotted lines indicate this for the hydrodynamical and dark matter only simulation respectively.} \label{fig:merger_halo9181}
\end{figure*}

\begin{figure*}
	\centering
	\includegraphics[width=0.65\textwidth]{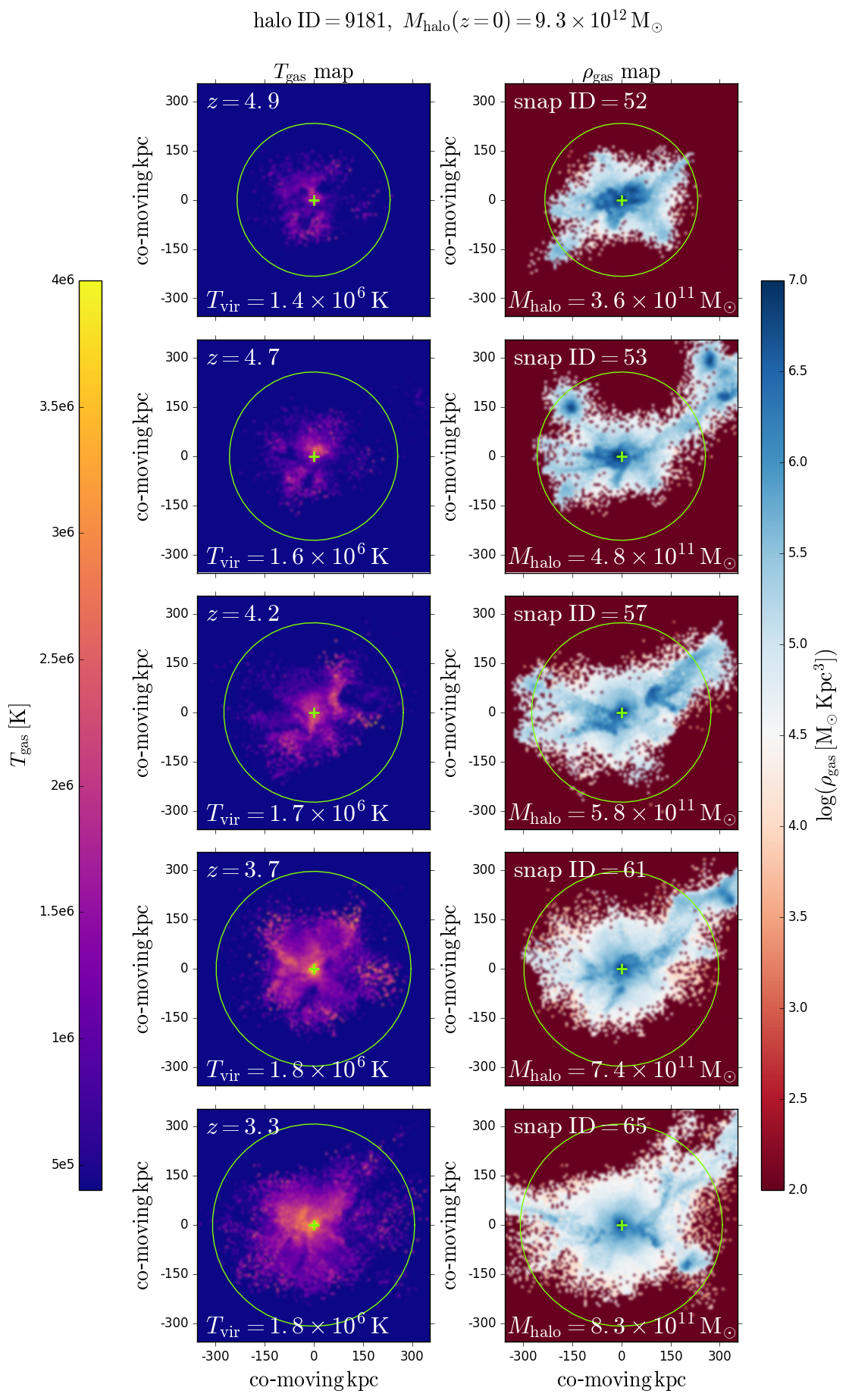}
	\caption{The projected temperature (left) and density (right) maps of the FOF group containing halo $9181$ for the selected snapshots labeled in the lower left panels (region A) of Fig.~\ref{fig:merger_halo9181}. The colour scale for temperature is linear, in units of $K$, while the scale for density is logarithmic, with density in units of $\Msol\,{\rm kpc}^{-3}$. In each panel, the green circle indicates $r_{\rm vir}$ of the Dhalo, while the green cross shows its centre. The snapshot ID, redshift, halo mass and $T_{\rm vir}$ for each snapshot are given in each panel. } \label{fig:merger_halo9181_details_A}
\end{figure*}

\begin{figure*}
	\centering
	\includegraphics[width=0.65\textwidth]{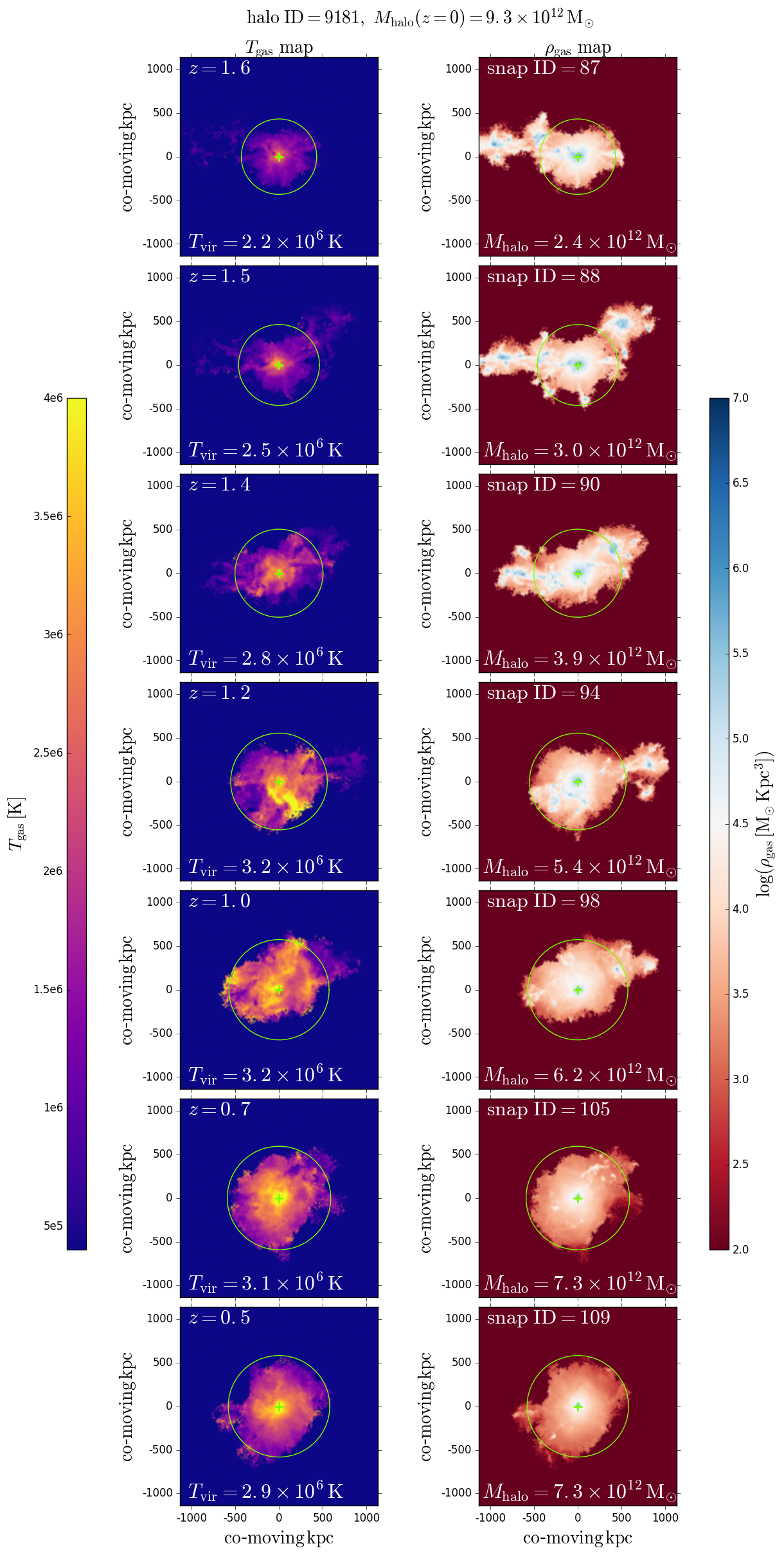}
	\caption{The projected temperature (left) and density (right) maps of the FOF group containing halo $9181$ for the selected snapshots labeled in the lower right panels (region B) of Fig.~\ref{fig:merger_halo9181}. The colour scales have the same meaning as in Fig.~\ref{fig:merger_halo9181_details_A}; see its caption for more information.  } \label{fig:merger_halo9181_details_B}
\end{figure*}

The above example is for a major merger that happens at low
redshift. Next we consider a halo major merger at high
redshift in a halo whose final mass is
    $8.3\times 10^{11}\Msol$, that is much smaller than in the
    previous example. For this purpose, we selected halo $9181$. It
experiences a halo merger with mass ratio about $3:1$ at $z\sim
5$. The mass cooling rate as a function of redshift for this halo is
shown in Fig.~\ref{fig:merger_halo9181}. The period corresponding to
this major merger is labeled `A' in the top panel and the zoom-in
plots are shown at the lower left. 

It seems that the hydrodynamical simulation does not predict any drop in the mass cooling rate related to this merger. The SA model predicts some drop, but the cooling rate is reduced by only about a factor $2$, which is much weaker than for the low redshift halo merger discussed above. Thus, neither the simulation nor the SA model predict a strong suppression of gas cooling for this major merger. To investigate the reason for this, the temperature and density maps of the gas were generated for snapshots just before the merger, just after the Dhalo merger and one, two and three halo dynamical timescales after the merger. These selected snapshots are labelled with magenta dots in the lower left panel of Fig.~\ref{fig:merger_halo9181}, while the corresponding maps are shown in Fig.~\ref{fig:merger_halo9181_details_A}.

From the density maps in Fig.~\ref{fig:merger_halo9181_details_A}, it
is clear that at this high redshift, the gas is filamentary rather
than in a spherical gas halo. However, the gas filaments can hardly be
seen in the temperature maps, because the colour scale for the
temperature is set to be sensitive
only to the gas with $T\gsim T_{\rm vir}$, while the
  filamentary gas is cold, with $T<T_{\rm vir}$
  \citep[c.f.][]{Nelson_IGM_sim,cold_accretion_keres}.

Just before the merger, at snapshot $52$, the halo gas is dominated by the filamentary cold gas, while the temperature map shows that the hot gas halo is less developed. Then, for the snapshots shown after the merger, the density maps continue to indicate the existence of filamentary gas, while a hot gas halo component becomes more and more obvious. There is no strong shock as in Fig.~\ref{fig:merger_halo4594_details} for the low redshift merger, and the development of the hot gas halo is more associated with the gradual transition from the filamentary accretion to the slow cooling regime, as the halo gradually grows in mass from $3\times 10^{11}\Msol$ to $8\times 10^{11}\Msol$, so it is largely unconnected to the major merger.

It is difficult for shocks to heat the cold filamentary gas. Therefore, a major merger hardly suppresses the cooling. In the SA model, there is no filamentary gas, but for this relatively low halo mass, the assumed hot gas halo is close to the fast cooling regime, in which the cooling timescale is very short, and so significant heating of the gas is also very difficult. Therefore, the SA model does not predict a strong suppression of gas cooling either and the predicted gas cooling rate is close to that in the hydrodynamical simulation.

Halo $9181$ shows a deep drop in the mass cooling rate at $z \approx 0.7$ in the hydrodynamical simulation, as shown in the region labeled `B' in Fig.~\ref{fig:merger_halo9181}. From the zoom-in plots in the lower right corner, it is seen that this drop is not caused by a single major merger, but by a series of smaller mergers. Two mergers with mass ratios $4:1$ and $5:1$ take place in quick succession, and together give rise to a rapid mass increase. The Dhalo merger completes at snapshot $90$, while from the density and temperature maps in Fig.~\ref{fig:merger_halo9181_details_B}, again the merged halo has not yet relaxed at this snapshot, and the relaxation happens during the following two halo dynamical timescales. The gaseous halo is heated up during this process, as can be seen from the temperature maps for snapshots $94$ and $98$.

After snapshot $90$, there are still some relatively rapid episodes of mass growth, and this, together with the heating induced by the two mergers, causes a deep drop in the cooling rate in the hydrodynamical simulation at snapshot $105$, and about one halo dynamical timescale later, at snapshot $109$, the cooling rate rises again. In the SA model, the reduction in the mass cooling rate happens immediately after the completion of the Dhalo merger.  Because of the absence of any time delay, the effects caused by the further mass increases after snapshot $94$ do not superpose onto the effects of these two mergers, and so only cause small ripples in the SA cooling rate following the deep dip (see green dashed ellipse in the lower right panel of Fig.~\ref{fig:merger_halo9181}). The deep drop in the SA cooling rate is weaker than that in the simulation.

Overall, we found that rapid gas accretion induced by halo mergers does suppress gas cooling, but this happens only for merger events at low redshifts and for halos in the slow cooling regime. Previously \citet{monaco_2014_comp} also investigated the suppression of cooling by major mergers. That work was based on  SPH simulations. \citeauthor{monaco_2014_comp} found no anti-correlation between the ratios of halo mass and of mass cooling rate for two adjacent snapshots. The cooling rates were taken either from the same snapshots from which the halo masses were taken, or from snapshots a few halo dynamical timescales later. This lack of correlation was interpreted as meaning that no systematic suppression of cooling due to halo major mergers was seen. From our results, this could be partially caused by the mixing of mergers at both high and low redshifts. \citet{monaco_2014_comp} also provided results of two individual mergers. These are at $z< 1$, but from Fig.11 of \citeauthor{monaco_2014_comp}, there still seems to be no strong drop in cooling rates. We noticed that in \citet{monaco_2014_comp} the mass cooling rate was measured for the entire FOF group, while here we measure this for each central galaxy, but we checked that these different ways of performing the measurements does not significantly weaken the drop, as shown by the blue dashed line in the left panel of Fig.~\ref{fig:merger_halo4594}. It is still possible that measuring the total mass cooling rate in the FOF group can mask the drop in cooling rate in some cases. The differences between our results and those of \citet{monaco_2014_comp} could also be caused by the differences between the SPH and moving mesh methods for hydrodynamical simulations.

\subsubsection{Artificial Effects}
\label{sec:results_artificial_effects}
In the left panel of Fig.~\ref{fig:cooling_results_fast_cooling1}, the mass cooling rate measured from the hydrodynamical simulation drops sharply at $z\sim 0$. This kind of phenomenon is mainly observed in halos with $M_{\rm halo}(z=0)<10^{12}\Msol$ (but not all halos in this mass range show this kind of drop). We checked that this is because for these relatively low mass halos, at $z\sim 0$, about $80\%$ of the total baryons have already cooled down and been turned into stars by our star formation recipe; the gas in the central galaxy becomes so diffuse that its density falls below the threshold for star formation. By only counting stars, we then left out this part of the cold gas. Also note that at $z\sim 0$, the remaining gas is typically accreted onto the central galaxy with higher angular momentum than the gas accreted at early times (because, on average, the angular momentum of the dark matter halo increases with mass growth), and this is another factor that reduces the gas density in the central galaxy. Since this effect of the cold gas density dropping below the star formation threshold only happens in some low mass halos at $z\sim 0$, omitting it does not strongly change our results for the cumulative cooled down mass or for the evolution of mass cooling rates over a large redshift range.

In the left panels of Fig.~\ref{fig:cooling_results_fast_cooling1} and \ref{fig:cooling_results_fast_cooling2}, it is also seen that the increase of mass cooling rates at high redshift in the simulation is more gradual and appears earlier than in the SA model. This is an artificial effect caused by our temperature threshold for cooling. According to Eq(\ref{eq:cooling_T_limit}), a gas cell is allowed to cool only if its temperature is high enough. The temperature threshold roughly corresponds to the virial temperature of a halo with mass $2\times 10^{10}\Msol$. In a halo with $M_{\rm halo}\ll 2\times 10^{10}\Msol$, an artificial hot gas halo forms due to the absence of radiative cooling. As shown by Fig.~\ref{fig:fit_hot_gas_profile}, in the simulation the hot gas in the central region of a halo tends to have higher temperature. Therefore, in the simulation, when a halo is still below $2\times 10^{10}\Msol$, the cooling has already begun in its central region, and later, when the halo is more massive, the cooling gradually extends over the whole hot gas halo. Therefore, the mass cooling rate gradually rises in the simulation. In the SA model, since it is assumed that the hot gas halo has a temperature equal to $T_{\rm vir}$ independently of radius, the hot gas halo can only start cooling when the halo mass reaches $2\times 10^{10}\Msol$, and once cooling is allowed, the whole hot gas halo starts cooling immediately. Therefore, the cooling rate rises sharply, but slightly later than in the simulation.

We also note that in some rare cases the SA model starts cooling
earlier than the simulation. This is because the Dhalo masses from the
  dark matter only simulations (used in the SA model) and
  hydrodynamical simulations can be slightly different, and sometimes
  this difference causes the halo in the SA model to pass the threshold
  for cooling first. 


\subsection{Comparison with different semi-analytical models: case studies}
\label{sec:results_model_comp_case_study}
\begin{figure*}
	\centering
	\includegraphics[width=0.7\textwidth]{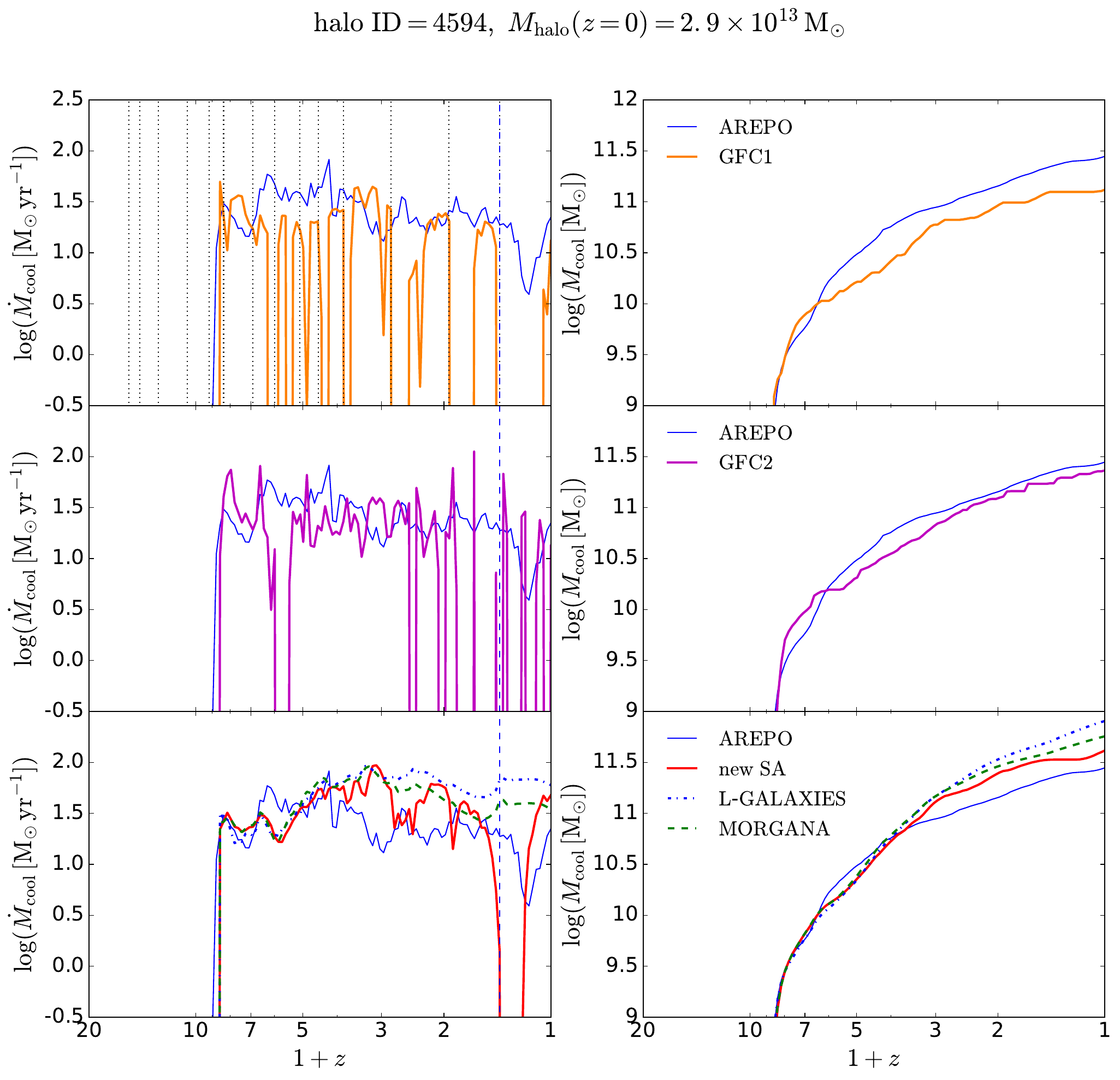}
	\caption{{\it Left column}: the mass cooling rates predicted by different SA models for halo $4594$ ($M_{\rm halo}(z=0)=2.9\times 10^{13}\Msol$), which enters the slow cooling regime at low redshift. In all panels, the blue solid line shows the mass cooling rate from the hydrodynamical simulation. This is compared with the \GALFORM GFC1 model in the top panel, with the \GALFORM GFC2 model in the middle panel, and with the \lgalaxy, \MORGANA and new cooling models in the bottom panel. The blue vertical dashed line in each panel indicates the redshift of a halo major merger, which causes the drop in cooling rate in both the hydrodynamical simulation and the new cooling model. Here this redshift is derived from the Dhalo merger tree used in the SA models, i.e.\ constructed from the dark matter only simulation. In the top panel, the vertical dotted lines indicate the artificial halo formation events calculated for the GFC1 cooling model. These formation events are not used in the other models, so these lines are omitted in the other panels. {\it Right column}: the cumulative cooled down mass predicted by the simulation and SA models. The line styles are the same as those in the left column.}
\label{fig:SA_comp_indi_halo_slow_cooling}
\end{figure*}

We compare the mass cooling rates predicted by different SA models
with the results from the hydrodynamical simulation for two halos
of very different masses that have already been
  studied in \S\ref{sec:results_cooling_physics}. In
Fig.~\ref{fig:SA_comp_indi_halo_slow_cooling} we compare mass cooling
rates for the high-mass halo $4594$, which has $M_{\rm halo}
=2.9\times 10^{13}\Msol$ at $z=0$, and enters the slow cooling regime
at low redshift. We first consider the GFC1 cooling model, which has
been widely used in different versions of \GALFORM. This is shown in
the top row of the figure. This model generates many sharp drops in
the cooling rate that do not appear in the simulation results. The
majority of these drops are seen to be associated with the artificial
halo formation events introduced in the GFC1 model (as described in
\S\ref{sec:method_SA_models_GFC1}), as many drops appear just after
the vertical dotted lines indicating the redshifts of these
events. The halo formation events cause drops in the mass cooling rate
because the time available for cooling, $t_{\rm cool,avail}$, is reset
to zero at each halo formation event, which means the whole hot gas halo ``forgets''
its previous cooling history and is effectively newly heated up. At
high redshifts, e.g.\ $z\geq 6$, the halo formation events only cause
small drops in the cooling rate, because the cooling timescale is very
short for high redshift, low mass halos, while at lower redshifts, the
cooling timescale becomes increasingly long, so that just after a halo
formation event the gas has to wait for a longer and longer time to
cool down. During this wait, the cooling rate drops to zero, and
correspondingly, wider and wider drops appear.

However, there are some drops in the cooling rate in the GFC1 model that are not associated with halo formation events. These drops are caused by the way in which $t_{\rm cool,avail}$ is estimated. At any given moment, the GFC1 model calculates a cooling radius, $r_{\rm cool}$, and assumes that all gas within $r_{\rm cool}$ has cooled down by this time. $r_{\rm cool}$ itself is determined from the condition $t_{\rm cool}(r_{\rm cool})=t_{\rm cool,avail}$, where $t_{\rm cool}(r)$ is the cooling timescale at radius $r$ for the current hot gas halo, while $t_{\rm cool,avail}$ is determined by the previous history of the halo. The GFC1 model assumes that $t_{\rm cool,avail}$ is always the physical time since the last halo formation event, which would be correct if the cooling started from the last halo formation event, and the hot gas halo remained fixed during cooling. However, the GFC1 cooling model allows the hot gas to evolve due to the growth of the dark matter halo. From high to low redshift, the halo evolves to acquire a gradually lower mean density, and this causes $t_{\rm cool}$ to gradually become longer. However, the estimation of $t_{\rm cool,avail}$ does not include a corresponding adjustment, so $t_{\rm cool,avail}$ is effectively underestimated, leading to an underestimation of $r_{\rm cool}$. This can cause $r_{\rm cool}$ to become smaller than the cooling radius at the previous time step, in which this model then determines that there is no cooling for the current time step, and there is a sharp drop in the mass cooling rate.

Also, in the GFC1 cooling model, the halo virial velocity, $V_{\rm vir}$, is only updated at each halo formation event (\S\ref{sec:method_SA_models_GFC1}). When the halo grows in mass between two halo formation events, its $V_{\rm vir}$ is kept unchanged, and is typically smaller than the virial velocity would be if it were calculated using the current halo mean density according to the spherical collapse model, $\Delta'_{\rm vir} \rho_{\rm crit}$. Conequently, $r_{\rm vir}$ calculated from this velocity is overestimated, while the mean halo density is underestimated. This artificial underestimation in density further worsens the underestimation of $t_{\rm cool,avail}$, and causes these drops to appear more frequently.

\begin{figure*}
	\centering
	\includegraphics[width=0.7\textwidth]{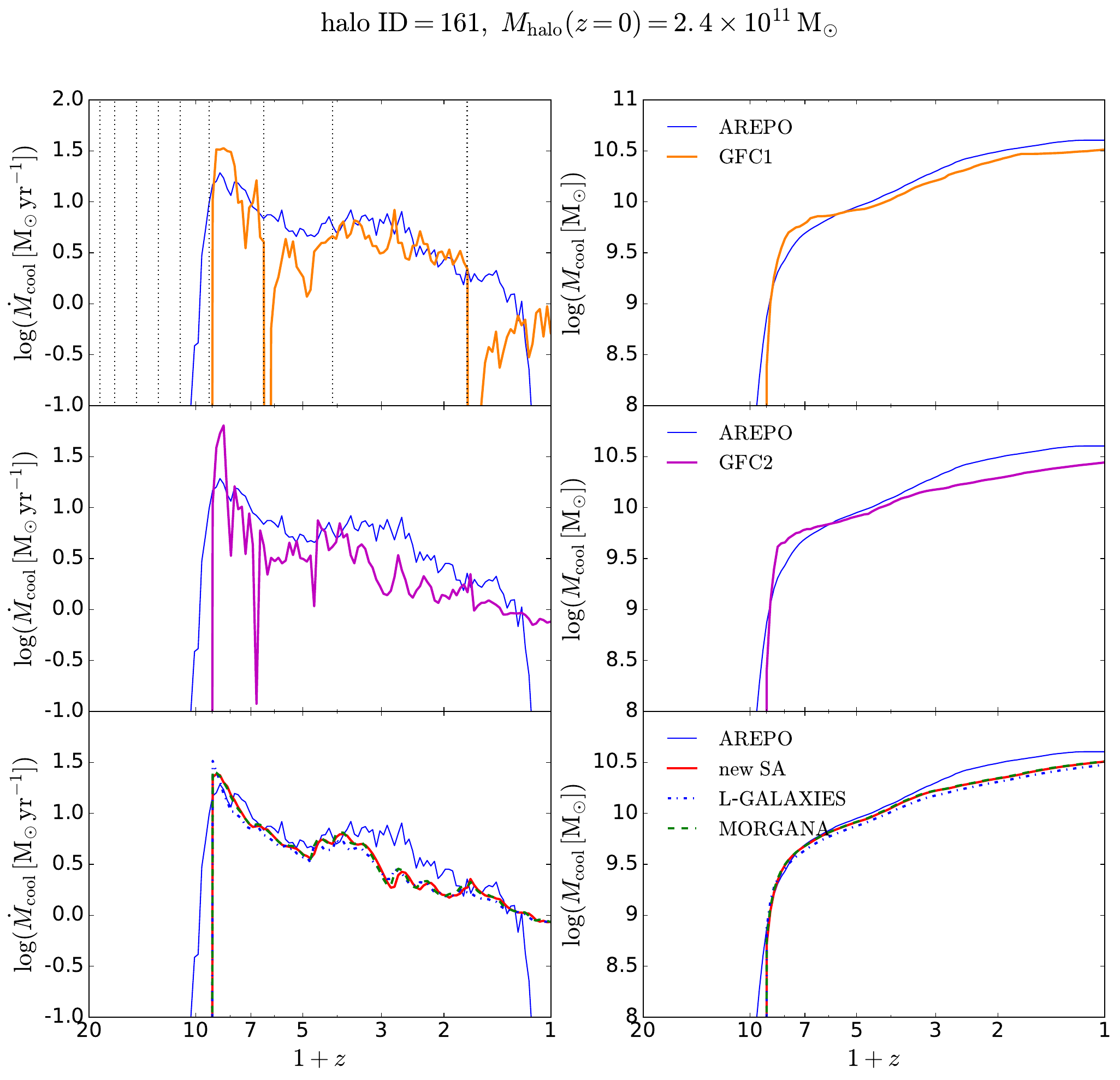}
	\caption{As Fig.~\ref{fig:SA_comp_indi_halo_slow_cooling}, but for halo $161$ ($M_{\rm halo}(z=0)=2.4\times 10^{11}\Msol$), which is in the fast cooling regime over the whole of its evolution.}
\label{fig:SA_comp_indi_halo_fast_cooling}
\end{figure*}

The GFC2 model aims to remove the dependence of gas cooling on the artificial halo formation events, to make the predicted cooling history more continuous. The middle row in Fig.~\ref{fig:SA_comp_indi_halo_slow_cooling} compares the results from the GFC2 cooling model with the hydrodynamical simulations. Surprisingly, although the formulation of this model is intended to make the cooling continuous, the actual predicted cooling history still shows many sharp drops. These drops are mainly caused by the very rough calculation of the total energy lost by cooling (equation~\ref{eq:E_cool_GFC2}) and by the method of calculating the time available for free-fall, $t_{\rm ff,avail}$, as we now explain.

Just as in the new cooling model, the GFC2 model accumulates the total energy radiated away as a record of the cooling history of the hot gas halo. Then, at any given time, this energy is divided by the cooling luminosity of the current hot halo to derive $t_{\rm cool,avail}$ for the current halo. This method includes the effects of the hot gas halo evolution on $t_{\rm cool,avail}$, so, in principle, it should avoid the problems identified in the GFC1 model above. However, because a very rough approximation is employed to calculate the cooling luminosity and the total energy radiated away, the effects of halo evolution in the calculations of these two quantities do not necessarily match those in the calculation of $t_{\rm cool}$, and sometimes this model still underestimates $r_{\rm cool}$. When $r_{\rm cool}$ for the current time step becomes smaller than that for the previous time step, the model again sets the mass cooling rate to zero.

Now we consider how the calculation of $t_{\rm ff,avail}$ causes drops in cooling rates in the GFC2 model. In both the GFC1 and GFC2 models, although it is assumed that the gas within $r_{\rm cool}$ has cooled down, this gas has not necessarily been accreted by the central galaxy, because it may not have had enough time to fall in under gravity. This effect is included in these two models through a free-fall radius, $r_{\rm ff}$, defined as $t_{\rm ff}(r_{\rm ff})=t_{\rm ff,avail}$, where $t_{\rm ff}(r)$ is the gravitational free-fall timescale for radius $r$, and $t_{\rm ff,avail}$ is the time available for free fall. Only the gas within both $r_{\rm cool}$ and $r_{\rm ff}$ is accreted by the central galaxy. The GFC2 model uses the same method to calculate $t_{\rm ff,avail}$ as for the calculation of $t_{\rm cool,avail}$, namely the total energy radiated away divided by the current cooling luminosity. The timescale $t_{\rm ff,avail}$ is not allowed to exceed $t_{\rm ff}(r_{\rm vir})$, and once $t_{\rm ff,avail}$ becomes larger than this limit, the total energy used to derive it is reset to $t_{\rm ff}(r_{\rm vir})\times L_{\rm cool}$, with $L_{\rm cool}$ the cooling luminosity of the current hot gas halo.

In the GFC2 model, the accumulation of the total energy lost by radiative cooling, $E_{\rm cool}$, (equation~\ref{eq:E_cool_GFC2}) is biased low. This results as follows. The mass of the cooled gas, $M_{\rm cooled}$, is gradually removed from the hot gas halo, which  allows this halo to contract towards the halo centre. Accordingly, the contribution of this removed gas to the total energy radiated away should also be removed. The GFC2 cooling model subtracts the total thermal energy of the removed gas from the total energy radiated away according to equation~(\ref{eq:E_cool_GFC2}). This subtraction would be correct if this cooling model correctly accumulated the energy radiated away, which, for each gas shell is $\tilde{\Lambda}(T_{\rm vir}) \rho_{\rm hot}^2(r)dV\Delta t$, where $\tilde{\Lambda}(T)$ is the cooling function, $\rho_{\rm hot}(r)$ is the density of the shell of radius $r$ and $dV$ its volume, while $\Delta t$ is the timestep. However, the GFC2 model instead uses the rough approximation $\tilde{\Lambda} \rho_{\rm hot}(r)\bar{\rho}_{\rm hot}dV\Delta t$, with $\bar{\rho}_{\rm hot}$ the mean density of the hot gas halo, so if the cooling happens in the inner region of the hot gas halo (typical in the slow cooling regime), where $\rho_{\rm hot}(r)>\bar{\rho}_{\rm hot}$, then this approximation undererestimates the energy lost by cooling, and the above subtraction removes more energy than necessary. This would lead to an underestimation of $t_{\rm cool,avail}$, and since $t_{\rm ff,avail}$ is calculated in a similar way, it is also underestimated. Furthermore, at early times, the cooling is so fast that the derived $t_{\rm ff,avail}$ can easily lead to $r_{\rm ff}>r_{\rm vir}$, so the total energy used to calculate $t_{\rm ff,avail}$ is frequently reset to its limit value described above, while the energy used for $t_{\rm cool,avail}$ gradually accumulates to larger values. As a result, $t_{\rm ff,avail}$ is more sensitive to the biased subtraction. At late times, the underestimation of $t_{\rm ff,avail}$ can lead to $r_{\rm ff}$ being too small, and sometimes $r_{\rm ff}<r_{\rm cool}$ even for halos in the slow cooling regime. If the value of $r_{\rm ff}$ at the current timestep is smaller than that at the previous step, then no cool gas is accreted by the central galaxy, and there is a drop in the cooling rate.

Note that although both the GFC1 and GFC2 cooling models generate many artificial drops in mass cooling rates, the effects on the cumulative mass cooled down are not very strong, as can be seen from the right column in Fig.~\ref{fig:SA_comp_indi_halo_slow_cooling}. This is because typically each drop only lasts for a short time.

We compare the mass cooling rates from the \lgalaxy, \MORGANA and new cooling models with the hydrodynamical simulation in the bottom row of Fig.~\ref{fig:SA_comp_indi_halo_slow_cooling}. It can be seen that the \lgalaxy cooling model gives a very smooth evolution of cooling rate, which is better than the results from the two cooling models considered above. However, this model predicts that during the low redshift halo major merger (indicated by the blue vertical dashed line), there is no suppression of gas cooling, and instead the cooling rate increases by about a factor of two. A suppression is actually seen in the simulation and also predicted in the new cooling model, as discussed in detail in \S\ref{sec:results_halo_mergers}.

The behaviour of the \lgalaxy cooling model during the halo major merger can be understood as follows: This cooling model assumes $t_{\rm cool,avail}=t_{\rm dyn}$. Note that $t_{\rm dyn}=r_{\rm vir}/V_{\rm vir}$ is independent of halo mass, but evolves with redshift. Consider that halo major mergers typically only happen over a short time duration, so the redshift change during the merger can be ignored. Therefore $t_{\rm cool,avail}$ in this model is almost unaffected by a major merger. The gas accreted through such a merger is effectively assigned the previous cooling history of the main halo. 
This leads to the absence of  any suppression of the mass cooling rate.
Next consider that $r_{\rm vir}\propto M_{\rm halo}^{1/3}$, $T_{\rm vir}\propto M_{\rm halo}^{2/3}$ (the proportionality factors are constants for a given redshift) and for massive halos, the cooled down mass is still a small fraction of the total baryon mass, so roughly $M_{\rm hot}\propto M_{\rm halo}$. The \lgalaxy model assumes the hot gas density profile to be a singular isothermal, i.e.\ $\rho_{\rm hot}(r)=M_{\rm hot}/(4\pi r_{\rm vir})r^{-2}$, and so, according to the above scaling relations, $\rho_{\rm hot}(r)\propto M_{\rm halo}^{2/3}/r^2$, and $T_{\rm vir}/\rho_{\rm hot}(r) \propto r^2$, independent of $M_{\rm halo}$ for a given redshift. A halo major merger increases $M_{\rm halo}$ by up to a factor two, and so $T_{\rm vir}$ increases by a factor smaller than two.  For massive halos with $T_{\rm vir} \gtrsim 10^6\,{\rm K}$, the cooling function $\tilde{\Lambda}(T_{\rm vir})$ only increases slightly for this temperature change. Therefore the cooling timescale, $t_{\rm cool}(r)\propto T_{\rm vir}/(\tilde{\Lambda} \rho_{\rm hot}) \propto r^2/\tilde{\Lambda}$, is largely unchanged during a major merger for a given radius $r$, and so the cooling radius, $r_{\rm cool}$, which is derived from $t_{\rm cool}(r_{\rm cool})=t_{\rm dyn}$, is also largely unchanged. For massive halos, typically $r_{\rm cool}<r_{\rm vir}$, and in this case, the \lgalaxy cooling model calculates the mass cooling rate as $\dot{M}_{\rm cool}=(M_{\rm hot}/t_{\rm dyn})  (r_{\rm cool}/r_{\rm vir})$. Now it is obvious that the increase of $M_{\rm hot}$ during a major merger dominates the change of cooling rate (because $r_{\rm vir}\propto M_{\rm halo}^{1/3}$, while $M_{\rm hot}\propto M_{\rm halo}$), and enhances it by a factor of about two.

The \MORGANA cooling model also predicts a very smooth cooling history, which is quite similar to that predicted by the \lgalaxy model. The \MORGANA model used here does not include the additional suppression of cooling during major mergers that was imposed in \citet{morgana1}, and in this case, this model always assumes that each shell of the hot gas halo contributes to the current mass cooling rate, with a contribution, $dm (\Delta t/t_{\rm cool}(r))$, where $dm$ is the mass of a shell, $r$ its radius and $t_{\rm cool}(r)$ its cooling timescale, while $\Delta t$ is the timestep. This calculation also gives a smooth evolution of cooling rate during a major merger. Although the \MORGANA model assumes a hot gas profile different from the \lgalaxy model, the analysis described above for the effect of halo mergers in the \lgalaxy model still largely applies to \MORGANA. Then we can see that $t_{\rm cool}(r)$ is largely unchanged, while $dm$ is increased due to the merger, so the cooling rate is enhanced during a merger. Of course, the gas cooling during a major merger can be suppressed by incorporating the additional condition in the \MORGANA model, but unlike in the new cooling model, here this requires additional parameters to identify major mergers and determine the suppression duration. Also, as discussed in \S\ref{sec:results_halo_mergers} (for halo $9181$), in the hydrodynamical simulation high redshift major mergers do not suppress cooling, while, on the other hand, a sequence of smaller mergers at low redshift can  suppress cooling, and these cannot be captured by a recipe that simply suppresses cooling during a halo major merger.

In Fig.~\ref{fig:SA_comp_indi_halo_fast_cooling}, we perform the same
comparison of mass cooling rates between different SA models and the
hydrodynamical simulation for the low-mass halo $161$ (with $M_{\rm
  halo}=2.4\times 10^{11}\Msol$ at $z=0$). This halo provides a comparison focused on the
fast cooling regime. The GFC1 and GFC2 models again generate
artificial sharp drops in the predicted mass cooling
rates. The reasons for these drops are the same as discussed
above. Again these drops in the cooling rates have little effect on
the cumulative cooled down mass. The new cooling
model and the \lgalaxy and \MORGANA models predict very similar
outcomes for this halo, in better agreement with the simulation
results than the GFC1 and GFC2 models.

\subsection{Comparison with different semi-analytical models: statistical comparison}
\label{sec:stat_comparison}
We use the hydrodynamical simulation in the $50\,\Mpc$ cube for this study. We divided halos into four samples according to their mass at $z=0$. Specifically, these samples cover the mass ranges $10^{11}\Msol\leq M_{\rm halo}<3\times 10^{11}\Msol$, $3\times 10^{11}\Msol\leq M_{\rm halo}<10^{12}\Msol$, $10^{12}\Msol\leq M_{\rm halo}<10^{13}\Msol$ and $M_{\rm halo}\geq10^{13}\Msol$. According to the criterion in \S\ref{sec:results_cold_vs_filamentary}, the first range corresponds to halos mainly in the fast cooling regime, while the third and fourth ranges correspond to halos going into the slow cooling regime at low redshift, and the second range is a transition region. There are $1086$, $462$, $200$ and $24$ halos in the four mass ranges respectively. 

\begin{figure*}
	\centering
	\includegraphics[width=0.7\textwidth]{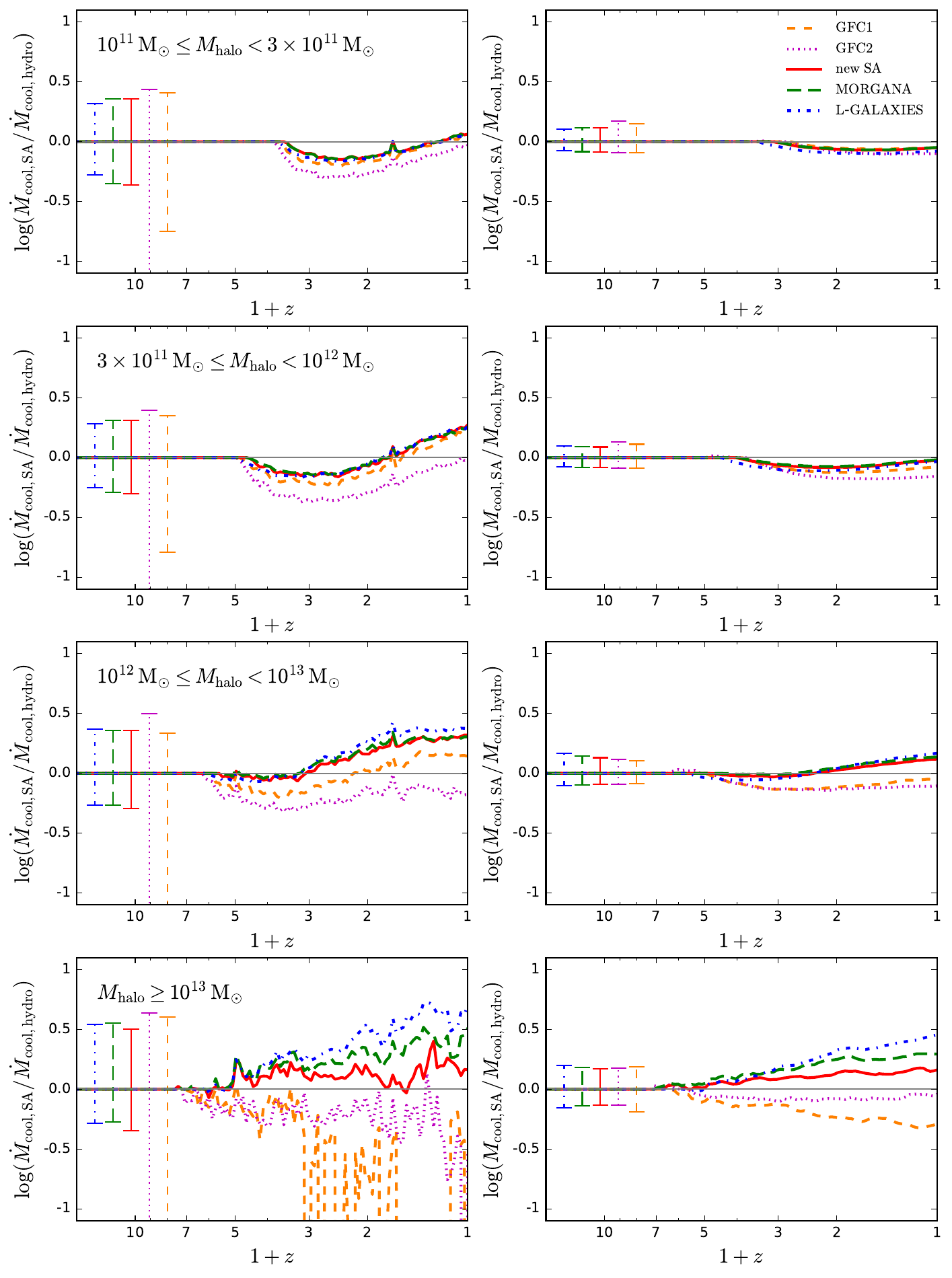}
	\caption{Statistical comparison between SA models and the hydrodynamical simulation. Each row corresponds to a different halo mass range, as labelled in each left-hand panel. The left panels show the mass cooling rate, $\dot{M}_{\rm cool}$, while the right panels show the cumulative mass cooled down, $M_{\rm cool}$. Each panel shows the logarithm of the ratio of the SA model prediction over the simulation prediction for the corresponding quantity, and the gray horizontal solid line indicates a ratio of one, i.e.\ the SA model and simulation giving the same prediction. Each line style corresponds to a different SA model, with the model name given in the key in the top right panel. The thick lines in each panel indicate the medians of the ratio, while the errorbars with corresponding line styles show the typical $10-90\%$ scatter around the medians.} 
\label{fig:SA_stat_comp}
\end{figure*}

Fig.~\ref{fig:SA_stat_comp} shows the medians and scatter of the individual halo differences between the SA models and the simulation, for the above-mentioned four halo mass ranges and for the cumulative mass that has cooled down, $M_{\rm cool}$, and the mass cooling rate, $\dot{M}_{\rm cool}$. In each panel, the thick lines show the medians of the differences, while the errorbars with corresponding line styles show the typical $10-90\%$ scatter around the medians.

From this figure it is clear that the different SA models all predict
cumulative cooled masses that are fairly close to the the simulation
results. The GFC1 and GFC2 models predict lower $M_{\rm cool}$ than
the simulation at low redshifts for halos with $M_{\rm
  halo}>10^{12}\Msol$. As analyzed in detail in \citet{new_cool}, this
is mainly caused by the lack of proper modelling of the hot halo
contraction. Without this contraction, the hot gas is always at
relatively low density, and so the cooling is slow because of the
strong dependence of the cooling luminosity on
density. Previously \citet{de_Lucia_2010_comp} had
  also found that compared to \lgalaxy and \MORGANA, the GFC1 model
  tends to underestimate cooling in massive halos, but instead of
  adding halo contraction, they suggested to adopt a steeper gas
  density profile, e.g.\ a singular isothermal, to mitigate this
  underestimation. The \lgalaxy and \MORGANA cooling models show
relatively large overestimations of $M_{\rm cool}$ for massive halos
($M_{\rm halo}\geq 10^{13}\Msol$). Overall, the new cooling model
agrees the best with the simulation over the four halo mass ranges.

In general terms, all of the SA cooling models also predict mass cooling rates, $\dot{M}_{\rm cool}$, that roughly agree with the simulation, but with larger scatter than the predictions for $M_{\rm cool}$. The typical $10\%$ scatter for the GFC1 and GFC2 models (shown by the lower boundaries of the corresponding errorbars) are much wider than for the other three models. The median for the GFC1 model in the mass range $M_{\rm halo}\geq10^{13}\Msol$ also shows a large deviation from the simulation results. Both of these results arise because these two models generate many artificial drops in the mass cooling rates (see Fig.~\ref{fig:SA_comp_indi_halo_slow_cooling} and discussions in \S\ref{sec:results_model_comp_case_study}), which lead to large underestimates compared to the simulation results.

Compared to the simulation, the mass cooling rates at $z>1$ of halos in the first two halo mass bins ($M_{\rm halo}(z=0)<10^{12}\Msol$) are underestimated by about $0.2\,{\rm dex}$ by all of the SA models. In the simulation, the central galaxies in these halos gain cold gas mainly through filamentary accretion (see \S\ref{sec:results_cold_vs_filamentary}), while in the SA models, these halos are mainly in the fast cooling regime. This underestimation indicates that although the mass accretion rates in both the filamentary accretion regime (in the simulation) and the fast cooling regime (in the SA models) are set roughly by the gravitational infall timescale, they are still slightly different from each other. Future direct modelling of the filamentary accretion in SA models may improve this agreement. 

At low redshifts, the new cooling model, \lgalaxy and the \MORGANA cooling model all tend to give higher cooling rates than the simulation. This is related to the potential overestimation of cooling by the SA models in the slow cooling regime, as discussed in \S\ref{sec:results_slow_cooling_regime}. The GFC1 and GFC2 models predict lower cooling rates than the other three models, for the reasons described above. 

The results from in this study are broadly consistent with the findings from the previous works mentioned in the Introduction, namely that at a rough level, different SA cooling models predict mass cooling rates and cumulative cooled down masses in agreement with hydrodynamical simulations. These previous comparisons were all based on SPH hydrodynamical simulations, and here we confirm this basic result using a state-of-the-art moving mesh hydrodynamical simulation. Some previous works \citep[e.g.][]{Lu_2011_comp, monaco_2014_comp} also noted that SA models in the fast cooling regime do not accurately describe the filamentary accretion seen in simulations, and found that SA models tend to underestimate the mass cooling rates in this regime. Here, using a different simulation technique, we qualititively confirm this point, but we find a smaller discrepancy between the SA models and the simulation. Some previous works \citep[e.g.][]{Saro_2010_comp, monaco_2014_comp} also noticed that compared to SPH simulations, SA models tend to overestimate cooling rates in high mass halos at low redshift. \citet{Saro_2010_comp} suggested that this discrepancy is caused by the hot gas halo density profile assumed in SA models, while our results suggest that this overestimation is at least partially caused by ignoring compression work in the SA models. \citet{monaco_2014_comp} also studied the effects of halo major mergers on cooling using SPH simulations, and concluded that mergers do not strongly effect mass cooling rates. In our work, as detailed in \S\ref{sec:results_halo_mergers} and \ref{sec:results_model_comp_case_study}, we uncovered a more complex picture for how halo mergers affect gas cooling, and found that under certain conditions halo mergers can strongly suppress cooling.


\section{summary} 
\label{sec:summary} 
In this work we have compared the gas cooling models from several
major semi-analytic (SA) galaxy formation models with simulations
performed using the state-of-the-art moving-mesh hydrodynamical code
\AREPO. All simulations and SA models have been run
  without any feedback or metal enrichment. The SA cooling models
considered here are the new \GALFORM cooling model introduced in
\citet{new_cool}, the GFC1 \citep{galform_bower2006} and GFC2
\citep{benson_bower_2010_cooling} models for \GALFORM, and the cooling
models from the \lgalaxy and \MORGANA models. The new cooling model is
possibly the most physically realistic model of the five. Our
comparison focuses on the total mass of gas that has cooled
down, $M_{\rm cool}$, and the mass cooling rate, $\dot{M}_{\rm cool}$.

Our comparison provides not only an assessment of the accuracy of each cooling model, but also some insights into the physics of gas cooling during cosmological structure formation. Our main conclusions are:
\begin{enumerate}[label=(\roman{*})]
\item For halos with $M_{\rm halo}\lsim 3\times 10^{11}\Msol$, the SA
  models predict the cooling to be in the fast cooling regime, in
  which the timescale for radiative cooling is shorter than or
  comparable to that for gravitational infall. However, the
  simulations show that for these halos the gas is mainly delivered to
  the central galaxy through cold filaments, and therefore the
  accretion is highly anisotropic. 
  Although these two pictures appear very different, the predicted
  mass accretion rates are similar, because in both pictures, these
  rates are mainly determined by the
      gravitational infall timescale.
	
\item In low redshift high mass halos, roughly spherical hot gas halos are seen in the simulations, in agreement with the slow cooling regime picture in the SA models. However, the simulations indicate that the gas maintains a roughly constant temperature around $T_{\rm vir}$ as it cools and contracts, until it reaches very high density, after which it cools down rapidly. This rough constancy of gas temperature is caused by the contraction work done by gravity when the gas gradually sinks towards the halo centre. During the entire cooling process, the total work done is about three times the initial gas thermal energy. The SA models typically do not consider this work, leading to overestimation of mass cooling rates in the slow cooling regime by a similar factor (even though the total cooled down masses in the SA models agree well with those in the simulations).
	
\item The simulations suggest that halo major mergers at low redshift can suppress cooling, while those at high redshift do not, because the cold filaments are hardly affected by mergers. At low redshift, a sequence of smaller mergers can also suppress cooling. The new cooling model better captures these effects of mergers compared to the other SA cooling models considered. This is an advantage of the new cooling model. The complex effects of halo mergers on cooling may explain the lack of correlation between the reduction of the cooling rate and the merger mass ratio reported in \citet{monaco_2014_comp}. \citeauthor{monaco_2014_comp} did not see any obvious suppression of cooling due to low redshift major mergers, and this may have been caused by the differences between the SPH hydrodynamical method used by them and the moving-mesh hydrodynamical method used in the present study.
	
\item Compared to the simulation results, all SA models give total cooled-down gas masses for halos with $M_{\rm halo}\geq10^{11}\Msol$ fairly close to the simulation results, with the GFC1 and GFC2 models tending to slightly underestimate this mass. The \lgalaxy and \MORGANA cooling models show a relatively large overestimation of the cooled masses for halos with $M_{\rm halo}\geq10^{13}\Msol$. The new cooling model agrees best with the simulations over the whole halo mass range.
	
\item All of the SA cooling models predict mass cooling rates that roughly agree with the simulations, but with scatter larger than that in the total cooled-down masses. The GFC1 and GFC2 models predict mass cooling rates significantly smaller than the simulation results in some cases. This is caused by artificial drops in cooling rates generated by the model formulations. In contrast, the new cooling model, as well as the \lgalaxy and \MORGANA models, predict more continuous evolution of the mass cooling rates. This is an advantage of these three models.
	
\end{enumerate}

In summary, it is reassuring that in spite of the simplifications required by SA models, a fundamental quantity such as the gas mass that cools during cosmological structure formation is quite accurately predicted, as judged by the results of a more general and precise full hydrodynamical simulation. This conclusion confirms the value and utility of semi-analytic modelling as a means to follow and quantify the process of galaxy formation.


\section*{Acknowledgements}
We thank Prof.\ Volker Springel for providing a specially modified version of the \AREPO code for our project. We thank Prof.\ Volker Springel, Dr. R$\ddot{{\rm u}}$diger Pakmor Dr. Sownak Bose and Dr. Yan Qu for their help in using \AREPO code. We also thank our referee Dr. Pierluigi Monaco for useful comments that helped to improve our paper. This work was supported by the Science and Technology Facilities Council [ST/L00075X/1] and European Research Council (ERC) Advanced Investigator Grant DMIDAS [GA 786910]. This work used the DiRAC Data Centric system at Durham University, operated by the Institute for Computational Cosmology on behalf of the STFC DiRAC HPC Facility (www.dirac.ac.uk. This equipment was funded by a BIS National E-infrastructure capital grant ST/K00042X/1, STFC capital grant ST/K00087X/1, DiRAC Operations grant ST/K003267/1 and Durham University. DiRAC is part of the National E-Infrastructure.

\bibliographystyle{mn2e}

\bibliography{paper_revised}

\appendix
\section{Determination of the core radius for the hot gas distribution}
\label{app:r_core_hot}
\begin{figure}
	\centering
	\includegraphics[width=0.45\textwidth]{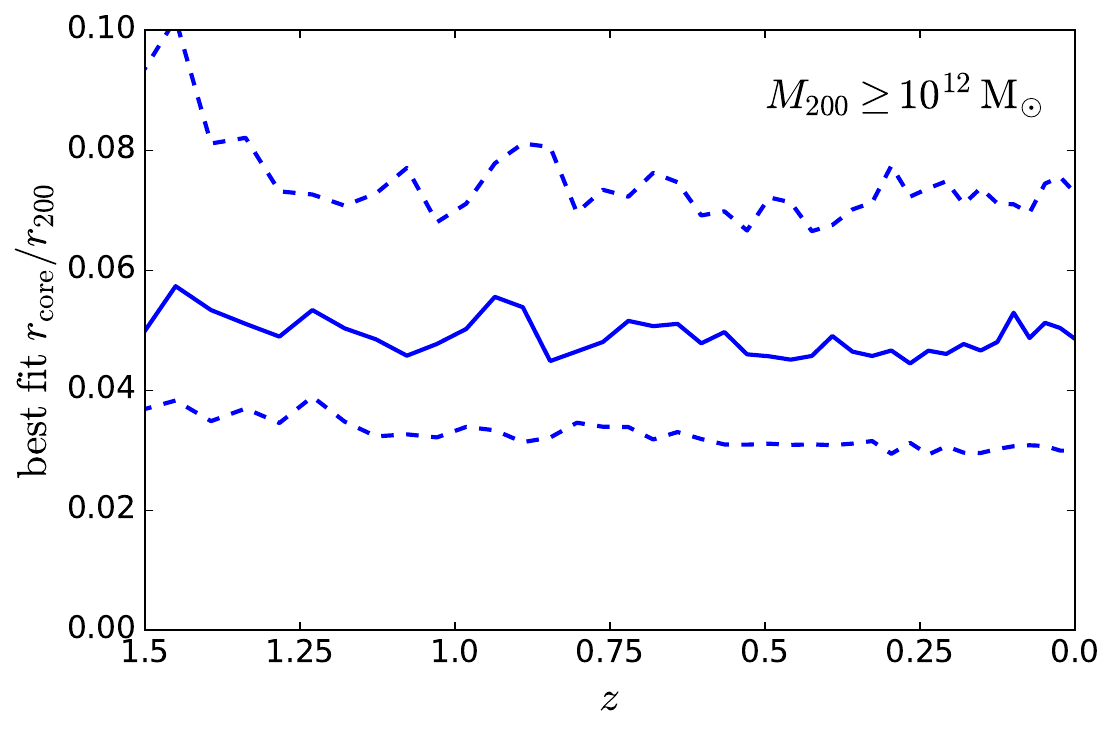}
	\caption{Best fit $r_{\rm core}/r_{200}$ for the gas in the most massive subgroups in the FOF groups with $M_{200}\geq 10^{12}\Msol$, for $z=0-1.5$. The blue solid line is the median of the best fit values, while the blue dashed lines indicate the $10-90\%$ range.} \label{fig:fit_hot_gas_profile}
\end{figure}

\begin{figure*}
	\centering
	\includegraphics[width=0.8\textwidth]{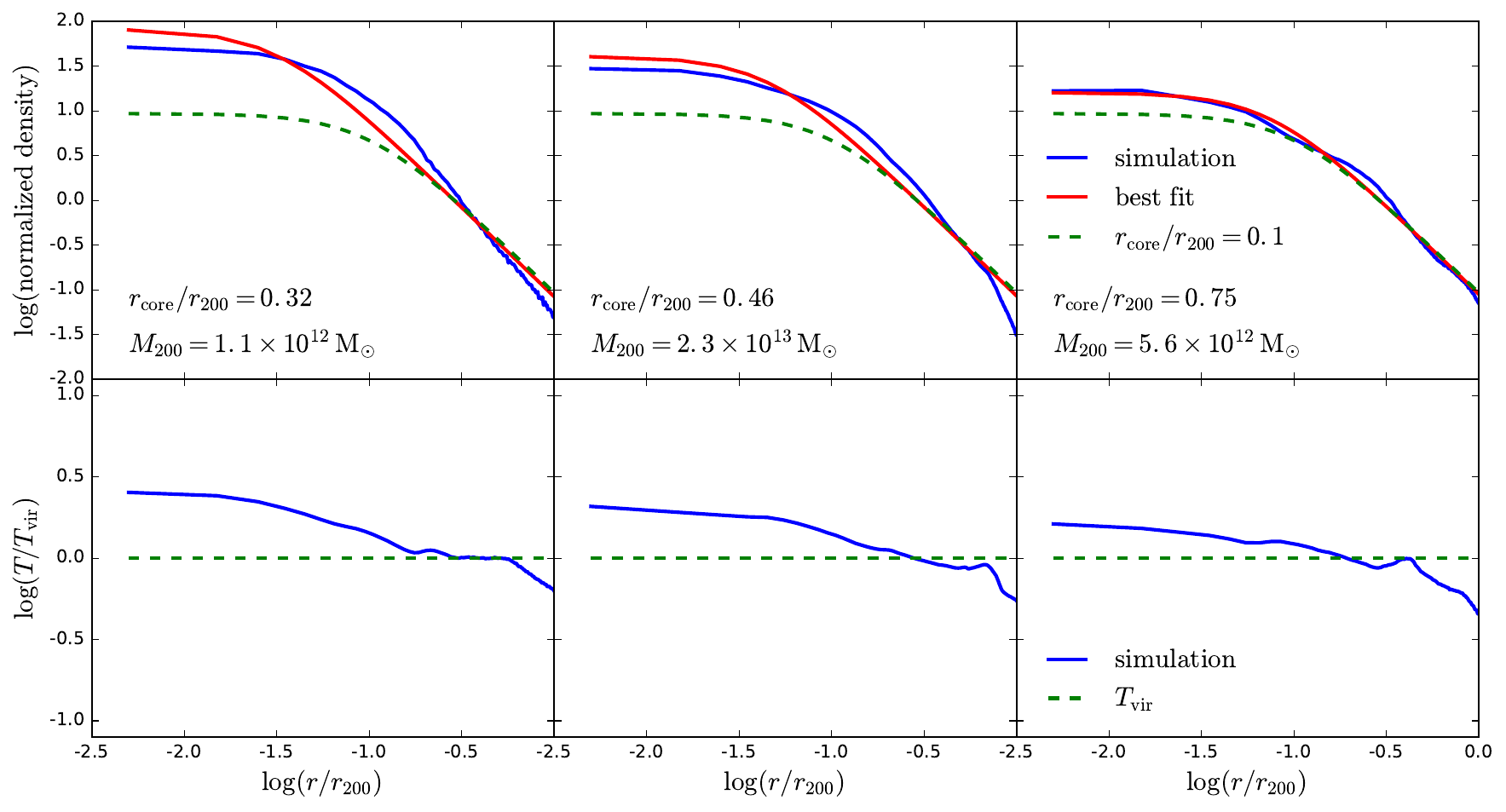}
	\caption{The spherical averaged gas density and temperature profiles of three example halos at $z=0$. Here the density profile is normalized so that the total mass within $r_{200}$ is $1$. The best-fit $\beta$-distribution density profile, as well as the profile with $r_{\rm core}/r_{200}=0.1$, are also shown.} \label{fig:fit_hot_gas_profile_example}
\end{figure*}

The GFC1, GFC2 and new SA cooling models all start from a hot gas halo with a density profile having a core radius,  $r_{\rm core}$.  To estimate what value of $r_{\rm core}$ to use in the SA models, we use results from the hydrodynamical simulation without radiative cooling described in \S\ref{sec:method_simulations}. We extract the spherically averaged density profile of the gas in the most massive subgroup in each of the FOF groups with $M_{200}\geq 10^{12}\Msol$, and then fit these profiles with the $\beta$-distribution (equation~\ref{eq:rho_hot_new_cool}), which has $r_{\rm core}$ as the only parameter, while its normalization is fixed by $r_{\rm core}$ and the total gas mass measured from the simulation. This fitting is performed at different redshifts, from $0$ to $1.5$. At $z=1.5$, there are $17$ FOF groups satisfying the selection condition, and this number gradually increases to $35$ FOF groups at $z=0$.

Fig.~\ref{fig:fit_hot_gas_profile} shows the median and $10-90\%$ range of the best fit $r_{\rm core}/r_{200}$ for different redshifts. It shows that $r_{\rm core}/r_{200}$ is very stable over redshift, with a median of about $0.05$. However, this value should not be directly used in SA cooling models for two reasons: 

Firstly, although the gas temperature profile is very shallow, it is not exactly constant. This can be seen from the three example halos shown in the bottom panels of Fig.~\ref{fig:fit_hot_gas_profile_example}. The temperature in the outer region ($r\geq 0.1r_{200}$) is close to $T_{\rm vir}$, but the central temperature is about a factor of two to three higher. Therefore, the constant temperature adopted by SA models, together with the best fit density profile, would lead to an overestimation of cooling in the SA models. In principle, we should adopt a more complex temperature profile in the SA models to solve this problem, but there is a simpler way to reduce this overestimation, which is by slightly increasing $r_{\rm core}$ to reduce the central density. As can be seen in Fig.~\ref{fig:fit_hot_gas_profile_example}, insofar as $r_{\rm core}\ll r_{200}$, this change only significantly affects the density in the central region, where the temperature is higher than $T_{\rm vir}$. Adopting $r_{\rm core}/r_{200}=0.1$ would lower the central density by a factor of two, which cancels the effect of underestimating the central temperature. We leave the application of non-constant temperature profiles in SA cooling models to future work.

Secondly, here the SA models adopt the Dhalo mass as the halo mass. The Dhalo mass is the sum of the subgroup masses in a given Dhalo, and is very close to the total mass of the given nonlinear structure. According to the spherical collapse model, the total mass corresponds to the virial mass, $M_{\rm vir}$, with associated radius,  $r_{\rm vir}$ and mean density, $\Delta'_{\rm vir}\rho_{\rm crit}$, where $\Delta'_{\rm vir}$ is the overdensity relative to $\rho_{\rm crit}$. On the other hand $r_{200}$ is associated with the mean density $200\rho_{\rm crit}$ and mass $M_{200}$. 

These two radii are related as $r_{200}/r_{\rm vir}=(M_{200}/M_{\rm vir})^{1/3}(\Delta'_{\rm vir}/200)^{1/3}$. We checked that the maximum difference between the Dhalo mass and $M_{200}$ is about $20\%$, so the departure of $r_{200}/r_{\rm vir}$ from unity is dominated by the second term, and therefore for the SA models we adopt
\begin{equation}
\frac{r_{\rm core}}{r_{\rm vir}}=\frac{r_{\rm core}}{r_{200}}\frac{r_{200}}{r_{\rm vir}}\approx 0.1\left( \frac{\Delta'_{\rm vir}}{200} \right)^{1/3} .
\end{equation}
We evaluate $\Delta'_{\rm vir}$ using the fitting formula \citep{spherical_over_density1,spherical_over_density2}:
\begin{equation}
\Delta'_{\rm vir}(z)=18\pi^2+82[\Omega_{\rm m}(z)-1]-39[\Omega_{\rm m}(z)-1]^2,
\end{equation}
where $\Omega_{\rm m}(z)$ is the matter density parameter at redshift
$z$. $\Delta'_{\rm vir}$ deviates significantly from $200$ only at
$z<1$, so $r_{\rm
  vir}$ and $r_{200}$ are significantly different only at
very low redshift. At $z=0$, this difference reaches its maximum, with
$r_{\rm vir}$ about $30\%$ larger than $r_{200}$.

\end{document}